\documentclass[twocolumn, preprint]{aastex631}

\shorttitle{Spectral Study of SS~433}
\shortauthors{HAWC Collaboration}
\graphicspath{{./}}

\begin{document}

\title{Spectral study of very high energy gamma rays from SS 433 with HAWC}

\correspondingauthor{C.D.~Rho}
\email{cdr397@skku.edu}
\correspondingauthor{Y.~Son}
\email{youngwan.son@cern.ch}
\correspondingauthor{K.~Fang}
\email{kefang@physics.wisc.edu}

\author{R.~Alfaro}
\affiliation{Instituto de F\'{i}sica, Universidad Nacional Aut\'{o}noma de M\'{e}xico, Ciudad de Mexico, Mexico} % IF-UNAM

\author[0000-0001-8310-4486]{C.~Alvarez}
\affiliation{Universidad Aut\'{o}noma de Chiapas, Tuxtla Guti\'{e}rrez, Chiapas, M\'{e}xico} % UNACH

\author{J.C.~Arteaga-Vel\'{a}zquez}
\affiliation{Universidad Michoacana de San Nicol\'{a}s de Hidalgo, Morelia, Mexico} % UMSNH

\author[0000-0002-4020-4142]{D.~Avila Rojas}
\affiliation{Instituto de F\'{i}sica, Universidad Nacional Aut\'{o}noma de M\'{e}xico, Ciudad de Mexico, Mexico} % IF-UNAM

\author[0000-0002-2084-5049]{H.A.~Ayala Solares}
\affiliation{Department of Physics, Pennsylvania State University, University Park, PA, USA} % PSU

\author[0000-0002-5529-6780]{R.~Babu}
\affiliation{Department of Physics and Astronomy, Michigan State University, East Lansing, MI, USA} % MSU

\author[0000-0003-3207-105X]{E.~Belmont-Moreno}
\affiliation{Instituto de F\'{i}sica, Universidad Nacional Aut\'{o}noma de M\'{e}xico, Ciudad de Mexico, Mexico} % IF-UNAM

\author{A.~Bernal}
\affiliation{Instituto de Astronom\'{i}a, Universidad Nacional Aut\'{o}noma de M\'{e}xico, Ciudad de Mexico, Mexico} % IA-UNAM

\author[0000-0002-4042-3855]{K.S.~Caballero-Mora}
\affiliation{Universidad Aut\'{o}noma de Chiapas, Tuxtla Guti\'{e}rrez, Chiapas, M\'{e}xico} % UNACH

\author[0000-0003-2158-2292]{T.~Capistr\'{a}n}
\affiliation{Instituto de Astronom\'{i}a, Universidad Nacional Aut\'{o}noma de M\'{e}xico, Ciudad de Mexico, Mexico} % IA-UNAM

\author[0000-0002-8553-3302]{A.~Carrami\~{n}ana}
\affiliation{Instituto Nacional de Astrof\'{i}sica, \'{o}ptica y Electr\'{o}nica, Puebla, Mexico} % INAOE

\author[0000-0002-6144-9122]{S.~Casanova}
\affiliation{Institute of Nuclear Physics Polish Academy of Sciences, PL-31342 IFJ-PAN, Krakow, Poland} % IFJ-PAN

\author[0000-0002-1132-871X]{J.~Cotzomi}
\affiliation{Facultad de Ciencias F\'{i}sico Matem\'{a}ticas, Benem\'{e}rita Universidad Aut\'{o}noma de Puebla, Puebla, Mexico} % FCFM-BUAP

\author[0000-0001-9643-4134]{E.~De la Fuente}
\affiliation{Departamento de F\'{i}sica, Centro Universitario de Ciencias Exactase Ingenierias, Universidad de Guadalajara, Guadalajara, Mexico} % UdG

\author[0000-0002-2672-4141]{D.~Depaoli}
\affiliation{Max-Planck Institute for Nuclear Physics, 69117 Heidelberg, Germany} % MPIK

\author{N.~Di Lalla}
\affiliation{Department of Physics, Stanford University: Stanford, CA 94305–4060, USA} % Stanford

\author[0000-0001-8487-0836]{R.~Diaz Hernandez}
\affiliation{Instituto Nacional de Astrof\'{i}sica, \'{o}ptica y Electr\'{o}nica, Puebla, Mexico} % INAOE

\author[0000-0001-8451-7450]{B.L.~Dingus}
\affiliation{Los Alamos National Laboratory, Los Alamos, NM, USA} % LANL

\author[0000-0002-2987-9691]{M.A.~DuVernois}
\affiliation{Department of Physics, University of Wisconsin-Madison, Madison, WI, USA} % UW-Madison

\author[0000-0001-5737-1820]{K.~Engel}
\affiliation{Department of Physics, University of Maryland, College Park, MD, USA} % UMD

\author[0000-0003-2423-4656]{T.~Ergin}
\affiliation{Department of Physics and Astronomy, Michigan State University, East Lansing, MI, USA} % MSU

\author[0000-0001-7074-1726]{C.~Espinoza}
\affiliation{Instituto de F\'{i}sica, Universidad Nacional Aut\'{o}noma de M\'{e}xico, Ciudad de Mexico, Mexico} % IF-UNAM

\author[0000-0002-8246-4751]{K.L.~Fan}
\affiliation{Department of Physics, University of Maryland, College Park, MD, USA} % UMD

\author[0000-0002-5387-8138]{K.~Fang}
\affiliation{Department of Physics, University of Wisconsin-Madison, Madison, WI, USA} % UW-Madison

\author[0000-0002-0173-6453]{N.~Fraija}
\affiliation{Instituto de Astronom\'{i}a, Universidad Nacional Aut\'{o}noma de M\'{e}xico, Ciudad de Mexico, Mexico} % IA-UNAM

\author{S.~Fraija}
\affiliation{Instituto de Astronom\'{i}a, Universidad Nacional Aut\'{o}noma de M\'{e}xico, Ciudad de Mexico, Mexico} % IA-UNAM

\author[0000-0002-4188-5584]{J.A.~Garc\'{i}a-Gonz\'{a}lez}
\affiliation{Tecnologico de Monterrey, Escuela de Ingenier\'{i}a y Ciencias, Ave. Eugenio Garza Sada 2501, Monterrey, N.L., Mexico, 64849} % ITESM

\author[0000-0002-1560-6334]{A.~Gonzalez Mu\~{n}oz}
\affiliation{Instituto de F\'{i}sica, Universidad Nacional Aut\'{o}noma de M\'{e}xico, Ciudad de Mexico, Mexico} % IF-UNAM

\author[0000-0002-5209-5641]{M.M.~Gonz\'{a}lez}
\affiliation{Instituto de Astronom\'{i}a, Universidad Nacional Aut\'{o}noma de M\'{e}xico, Ciudad de Mexico, Mexico} % IA-UNAM

\author[0000-0002-9790-1299]{J.A.~Goodman}
\affiliation{Department of Physics, University of Maryland, College Park, MD, USA} % UMD

\author{S.~Groetsch}
\affiliation{Department of Physics, Michigan Technological University, Houghton, MI, USA} % MTU

\author[0000-0001-9844-2648]{J.P.~Harding}
\affiliation{Los Alamos National Laboratory, Los Alamos, NM, USA} % LANL

\author[0000-0002-2565-8365]{S.~Hern\'{a}ndez-Cadena}
\affiliation{Tsung-Dao Lee Institute \& School of Physics and Astronomy, Shanghai Jiao Tong University, Shanghai, China} %SJTU

\author[0000-0001-5169-723X]{I.~Herzog}
\affiliation{Department of Physics and Astronomy, Michigan State University, East Lansing, MI, USA} % MSU

\author[0000-0002-5447-1786]{D.~Huang}
\affiliation{Department of Physics, University of Maryland, College Park, MD, USA} % UMD

\author[0000-0002-5527-7141]{F.~Hueyotl-Zahuantitla}
\affiliation{Universidad Aut\'{o}noma de Chiapas, Tuxtla Guti\'{e}rrez, Chiapas, M\'{e}xico} % UNACH

\author[0000-0002-3302-7897]{P.~H\"{u}ntemeyer}
\affiliation{Department of Physics, Michigan Technological University, Houghton, MI, USA} % MTU

\author[0000-0001-5811-5167]{A.~Iriarte}
\affiliation{Instituto de Astronom\'{i}a, Universidad Nacional Aut\'{o}noma de M\'{e}xico, Ciudad de Mexico, Mexico} % IA-UNAM

\author{S.~Kaufmann}
\affiliation{Universidad Politecnica de Pachuca, Pachuca, Hgo, Mexico} % UPP

\author[0000-0001-6336-5291]{A.~Lara}
\affiliation{Instituto de Geof\'{i}sica, Universidad Nacional Aut\'{o}noma de M\'{e}xico, Ciudad de Mexico, Mexico} % IGeof-UNAM

\author[0000-0002-2467-5673]{W.H.~Lee}
\affiliation{Instituto de Astronom\'{i}a, Universidad Nacional Aut\'{o}noma de M\'{e}xico, Ciudad de Mexico, Mexico} % IA-UNAM

\author[0000-0002-2153-1519]{J.~Lee}
\affiliation{University of Seoul, Seoul, Rep. of Korea} % UOS

\author[0000-0003-0513-3841]{C.~de Le\'{o}n}
\affiliation{Universidad Michoacana de San Nicol\'{a}s de Hidalgo, Morelia, Mexico} % UMSNH

\author[0000-0001-5516-4975]{H.~Le\'{o}n Vargas}
\affiliation{Instituto de F\'{i}sica, Universidad Nacional Aut\'{o}noma de M\'{e}xico, Ciudad de Mexico, Mexico} % IF-UNAM

\author[0000-0001-8825-3624]{A.L.~Longinotti}
\affiliation{Instituto de Astronom\'{i}a, Universidad Nacional Aut\'{o}noma de M\'{e}xico, Ciudad de Mexico, Mexico} % IA-UNAM

\author[0000-0003-2810-4867]{G.~Luis-Raya}
\affiliation{Universidad Politecnica de Pachuca, Pachuca, Hgo, Mexico} % UPP

\author[0000-0001-8088-400X]{K.~Malone}
\affiliation{Los Alamos National Laboratory, Los Alamos, NM, USA} % LANL

\author[0000-0002-2824-3544]{J.~Mart\'{i}nez-Castro}
\affiliation{Centro de Investigaci\'{o}n en Computaci\'{o}n, Instituto Polit\'{e}cnico Nacional, M\'{e}xico City, M\'{e}xico.} % CIC-IPN

\author[0000-0002-2610-863X]{J.A.~Matthews}
\affiliation{Dept of Physics and Astronomy, University of New Mexico, Albuquerque, NM, USA} % UNM

\author[0000-0002-8390-9011]{P.~Miranda-Romagnoli}
\affiliation{Universidad Aut\'{o}noma del Estado de Hidalgo, Pachuca, Mexico} % UAEH

\author{J.A.~Montes}
\affiliation{Instituto de Astronom\'{i}a, Universidad Nacional Aut\'{o}noma de M\'{e}xico, Ciudad de Mexico, Mexico} % IA-UNAM

\author[0000-0002-1114-2640]{E.~Moreno}
\affiliation{Facultad de Ciencias F\'{i}sico Matem\'{a}ticas, Benem\'{e}rita Universidad Aut\'{o}noma de Puebla, Puebla, Mexico} % FCFM-BUAP

\author[0000-0002-7675-4656]{M.~Mostaf\'{a}}
\affiliation{Department of Physics, College of Science \& Technology, Temple University, Philadelphia, PA 19122, USA} % Temple

\author[0000-0003-1059-8731]{L.~Nellen}
\affiliation{Instituto de Ciencias Nucleares, Universidad Nacional Aut\'{o}noma de Mexico, Ciudad de Mexico, Mexico} % ICN-UNAM

\author[0000-0002-6859-3944]{M.U.~Nisa}
\affiliation{Department of Physics and Astronomy, Michigan State University, East Lansing, MI, USA} % MSU

\author[0000-0001-7099-108X]{R.~Noriega-Papaqui}
\affiliation{Universidad Aut\'{o}noma del Estado de Hidalgo, Pachuca, Mexico} % UAEH

\author{Y.~P\'{e}rez Araujo}
\affiliation{Instituto de F\'{i}sica, Universidad Nacional Aut\'{o}noma de M\'{e}xico, Ciudad de Mexico, Mexico} % IF-UNAM

\author[0000-0001-5998-4938]{E.G.~P\'{e}rez-P\'{e}rez}
\affiliation{Universidad Politecnica de Pachuca, Pachuca, Hgo, Mexico} % UPP

\author[0000-0002-6524-9769]{C.D.~Rho}
\affiliation{Department of Physics, Sungkyunkwan University, Suwon 16419, South Korea} % SKKU

\author[0000-0003-1327-0838]{D.~Rosa-Gonz\'{a}lez}
\affiliation{Instituto Nacional de Astrof\'{i}sica, \'{o}ptica y Electr\'{o}nica, Puebla, Mexico} % INAOE

\author[0000-0001-6939-7825]{E.~Ruiz-Velasco}
\affiliation{Max-Planck Institute for Nuclear Physics, 69117 Heidelberg, Germany} % MPIK

\author[0000-0003-4556-7302]{H.~Salazar}
\affiliation{Facultad de Ciencias F\'{i}sico Matem\'{a}ticas, Benem\'{e}rita Universidad Aut\'{o}noma de Puebla, Puebla, Mexico} % FCFM-BUAP

\author[0000-0001-6079-2722]{A.~Sandoval}
\affiliation{Instituto de F\'{i}sica, Universidad Nacional Aut\'{o}noma de M\'{e}xico, Ciudad de Mexico, Mexico} % IF-UNAM

\author[0000-0001-8644-4734]{M.~Schneider}
\affiliation{Department of Physics, University of Maryland, College Park, MD, USA} % UMD

\author{J.~Serna-Franco}
\affiliation{Instituto de F\'{i}sica, Universidad Nacional Aut\'{o}noma de M\'{e}xico, Ciudad de Mexico, Mexico} % IF-UNAM

\author[0000-0002-1012-0431]{A.J.~Smith}
\affiliation{Department of Physics, University of Maryland, College Park, MD, USA} % UMD

\author[0000-0002-7214-8480]{Y.~Son}
\affiliation{University of Seoul, Seoul, Rep. of Korea} % UOS

\author[0000-0002-1492-0380]{R.W.~Springer}
\affiliation{Department of Physics and Astronomy, University of Utah, Salt Lake City, UT, USA} % University of Utah

\author[0000-0002-9074-0584]{O.~Tibolla}
\affiliation{Universidad Politecnica de Pachuca, Pachuca, Hgo, Mexico} % UPP

\author[0000-0001-9725-1479]{K.~Tollefson}
\affiliation{Department of Physics and Astronomy, Michigan State University, East Lansing, MI, USA} % MSU

\author[0000-0002-1689-3945]{I.~Torres}
\affiliation{Instituto Nacional de Astrof\'{i}sica, \'{o}ptica y Electr\'{o}nica, Puebla, Mexico} % INAOE

\author[0000-0002-7102-3352]{R.~Torres-Escobedo}
\affiliation{Tsung-Dao Lee Institute \& School of Physics and Astronomy, Shanghai Jiao Tong University, Shanghai, China} %SJTU

\author[0000-0003-1068-6707]{R.~Turner}
\affiliation{Department of Physics, Michigan Technological University, Houghton, MI, USA} % MTU

\author[0000-0002-2748-2527]{F.~Ure\~{n}a-Mena}
\affiliation{Instituto Nacional de Astrof\'{i}sica, \'{o}ptica y Electr\'{o}nica, Puebla, Mexico} % INAOE

\author[0000-0003-0715-7513]{E.~Varela}
\affiliation{Facultad de Ciencias F\'{i}sico Matem\'{a}ticas, Benem\'{e}rita Universidad Aut\'{o}noma de Puebla, Puebla, Mexico} % FCFM-BUAP

\author[0000-0001-6876-2800]{L.~Villase\~{n}or}
\affiliation{Facultad de Ciencias F\'{i}sico Matem\'{a}ticas, Benem\'{e}rita Universidad Aut\'{o}noma de Puebla, Puebla, Mexico} % FCFM-BUAP

\author[0000-0001-6798-353X]{X.~Wang}
\affiliation{Department of Physics, Michigan Technological University, Houghton, MI, USA} % MTU

\author{Z.~Wang}
\affiliation{Department of Physics, University of Maryland, College Park, MD, USA} % UMD

\author[0000-0003-2141-3413]{I.J.~Watson}
\affiliation{University of Seoul, Seoul, Rep. of Korea} % UOS

\author[0009-0006-3520-3993]{S.~Yu}
\affiliation{Department of Physics, Pennsylvania State University, University Park, PA, USA} % PSU

\author[0000-0002-9307-0133]{S.~Yun-C\'{a}rcamo}
\affiliation{Department of Physics, University of Maryland, College Park, MD, USA} % UMD

\author[0000-0003-0513-3841]{H.~Zhou}
\affiliation{Tsung-Dao Lee Institute \& School of Physics and Astronomy, Shanghai Jiao Tong University, Shanghai, China} %SJTU

\collaboration{82}{(HAWC collaboration)}
\noaffiliation

\begin{abstract}
Very-high-energy (0.1-100~TeV) gamma-ray emission was observed in HAWC data from the lobes of the microquasar SS~433, making them the first set of astrophysical jets that were resolved at TeV energies  \citep{Abeysekara:2018qtj}. In this work, we update the analysis of SS~433 using 2,565 days of data from the High Altitude Water Cherenkov (HAWC) observatory. Our analysis reports the detection of a point-like source in the east lobe at a significance of $6.6\,\sigma$ and in the west lobe at a significance of $8.2\,\sigma$. 
For each jet lobe, we localize the gamma-ray emission and identify a best-fit position. The locations are close to the X-ray emission sites ``e1'' and ``w1'' for the east and west lobes, respectively.
We analyze the spectral energy distributions and find that the energy spectra of the lobes are consistent with a simple power-law $\text{d}N/\text{d}E\propto E^{\alpha}$ with $\alpha = -2.44^{+0.13+0.04}_{-0.12-0.04}$ and $\alpha = -2.35^{+0.12+0.03}_{-0.11-0.03}$ for the east and west lobes, respectively. The maximum energy of photons from the east and west lobes reaches 56~TeV and 123~TeV, respectively. We compare our observations to various models and conclude that the very-high-energy gamma-ray emission can be produced by a population of electrons that were efficiently accelerated.

\end{abstract}


\submitjournal{ApJ}

%\received{11 June 2024}

%\revised{14 September 2024}

%\accepted{21 September 2024}

\section{Introduction} \label{sec:introduction}

SS~433, first discovered in 1977 \citep{1977ApJS.33.459S}, is a microquasar consisting of a compact object (i.e. a black hole or neutron star) that accretes matter from its massive companion star \citep{2004ASPRv..12....1F}. SS~433 is well-known for the emission from the jet interaction regions (also referred to as jet lobes) for which the jets terminate inside a nebula, W50, $\sim 40$~pc away from the binary system \citep{Safi-Harb_1997}.
The two jet emission regions, named ``east'' and ``west'', have been studied in depth over the decades in radio \citep{1980A&A....84..237G}, X-rays \citep{refId0,Safi-Harb_1999}, and gamma rays \citep{Abeysekara:2018qtj,Fang_2020,HESS24}. 

In 2018, the High Altitude Water Cherenkov (HAWC) observatory detected two multi-TeV gamma-ray hotspots spatially coincident with the known locations of the X-ray emission regions of SS~433 \citep{Abeysekara:2018qtj}. This discovery rekindled a series of observational campaigns that subsequently produced new results from experiments including the {\it Fermi} Large Area Telescope ({\it Fermi}-LAT; \citealp{Fang_2020}), Nuclear Spectroscopic Telescope Array (NuSTAR; \citealp{Safi-Harb_2022}), and High Energy Stereoscopic System (H.E.S.S.; \citealp{HESS24}). These results have provided further information about the TeV emission regions of SS~433.  

\begin{figure*}[ht!]
\centering
\gridline{\fig{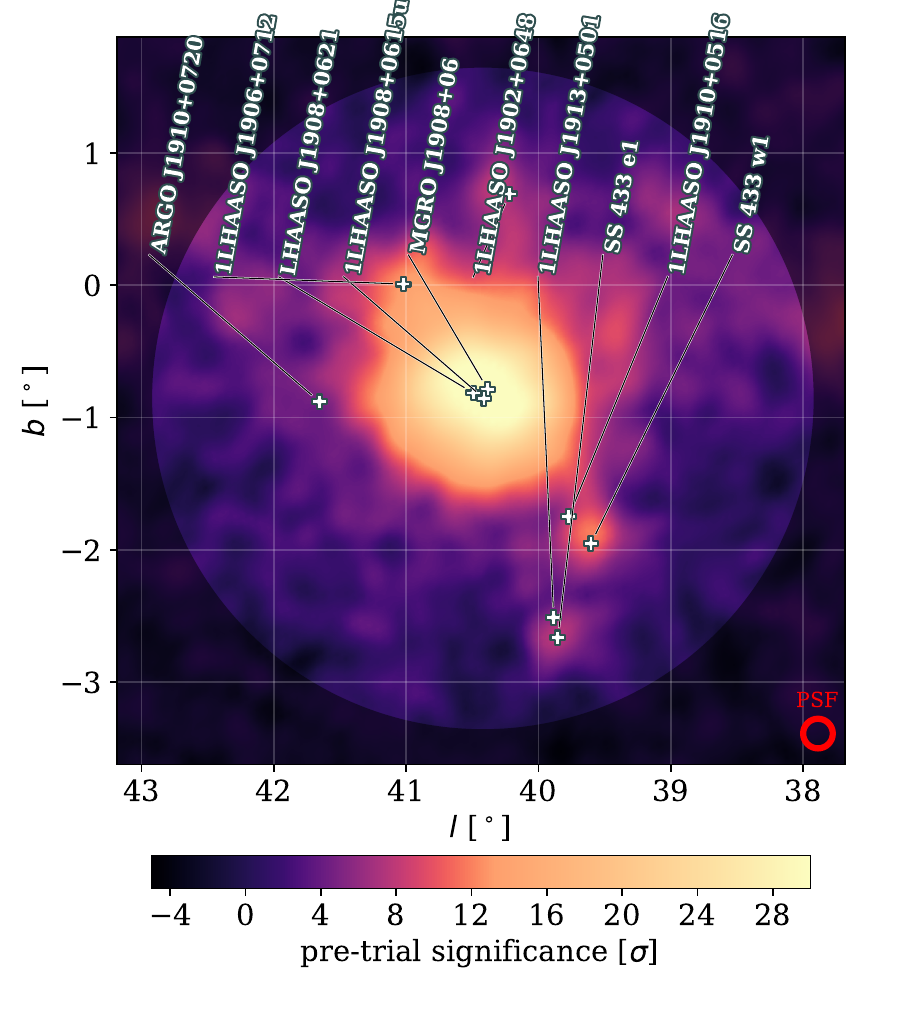}{\columnwidth}{(a)} \fig{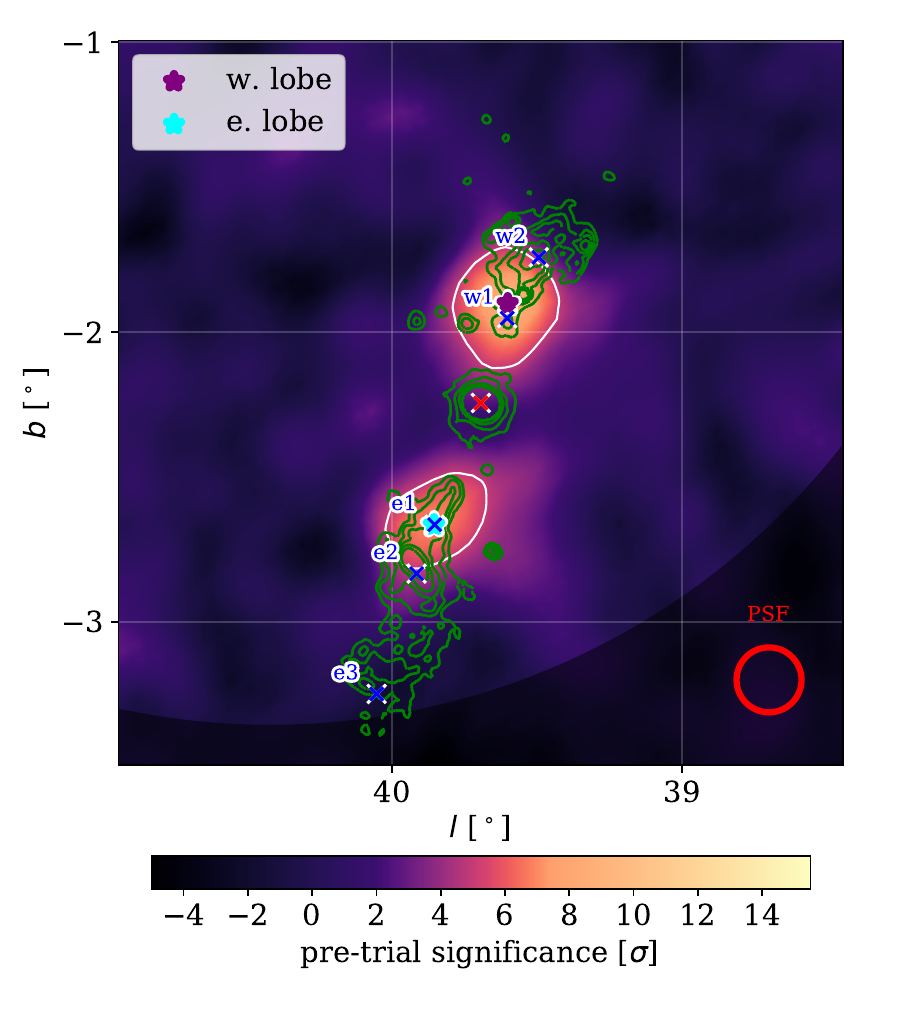}{\columnwidth}{(b)}}
\caption{
Significance maps above $1~\mathrm{TeV}$.
(a) A pre-trial significance map of the SS~433 region with labels from \citet{Abdo_2007, 2013ApJ...779...27B, Abeysekara:2018qtj, UHELHAASO, 1LHAASOCat}.
(b) A pre-trial significance map zoomed in closely around SS~433. The map shows excess after the model fit but only removing background sources for the visualization of the SS~433 jet lobes. The green contour lines indicate the observational results from ROSAT \citep{1996A&A...312..306B} and the white contour lines are the $5\sigma$ contour lines as observed by this work. The X-ray emission regions ``e1'', ``e2'', and ``e3'' for the east lobe and ``w1'' and ``w2'' for the west lobe have also been indicated \citep{Safi-Harb_1997}. The red cross marks the known location of the binary SS~433. The purple and cyan stars indicate the best-fit locations from this work. The red ring in each panel indicates the angular resolution. \label{fig:sigmap}}
\end{figure*} 

In this work, we present the updated results on the two jet lobes of SS~433. HAWC has accumulated $\sim1500$~days of more data compared to \citet{Abeysekara:2018qtj}. With 2,565~days of data and improved data reconstruction, we perform morphological study in the region-of-interest (ROI) \citep{Pass5Crab}. We identify significant levels of gamma-ray emission corresponding to the east lobe and the west lobe through a blind search. The ROI is the brighter region as shown in Figure~\ref{fig:sigmap}~(a).

In Section~\ref{sec:instrumentanddataanalysis}, we describe the HAWC observatory and the dataset used for the analysis. Also, we explain the source search pipeline used to identify the 6 sources found in the ROI through a blind search. In Section~\ref{sec:results}, we present the best-fit results on the SS~433 lobes and other sources found from the source search pipeline. We also show the spectra of the lobes computed with the HAWC data and the measured energy range of the observed gamma-ray photons. Finally, in Section~\ref{sec:conclusion}, we discuss the multi-wavelength modeling of the two lobes and comparisons with the results from \citet{Abeysekara:2018qtj}.

\section{Data and Analysis Method} \label{sec:instrumentanddataanalysis}
\subsection{Instrument and Data}
HAWC is a particle sampling array designed to observe very-high-energy (VHE) gamma-ray events with energies between hundreds of GeV and hundreds of TeV. HAWC is located at an altitude of 4,100~m next to Pico de Orizaba in Puebla, Mexico. The main array of HAWC consists of 300 water Cherenkov detectors (WCDs) providing a total geometrical area of $\sim220,000$~m$^2$ for indirect gamma-ray observations. Each WCD consists of a light-tight bladder containing $\sim200,000$~liters of purified water and four photomultiplier tubes anchored to the bottom. A WCD has a dimension of 7.3~m in diameter and 4.5~m in height that provides enough depth for the WCD to act as a calorimeter and a medium to allow air shower particles to emit Cherenkov light. 
HAWC has high operation time of over 95\% duty cycle and a wide simultaneous field of view of 2~sr. Typically, HAWC has an event rate of $\sim 25$~kHz of which $\sim0.1\%$ are gamma-ray events. More detailed information on HAWC and its operation can be found in \citet{2017ApJ...843...39A, HAWCDetector}.

Every gamma-ray event that HAWC detects is classified into an ``energy bin'' associated with a certain range of estimated energies \citep{Abeysekara:2019edl}. Currently, HAWC has two energy estimation algorithms, the ground parameter (GP) and neural network (NN) energy estimators. The GP algorithm uses the charge density at a fixed distance away from the axis of the gamma-ray air shower, while the NN algorithm uses a multilayer-perceptron architecture that employs parameters that are also used for HAWC event reconstructions \citep{Abeysekara:2019edl}. Note that the GP and NN datasets are comparable within their uncertainties \citep{Abeysekara:2019edl}. In this work, we have used the NN energy estimator binning scheme.

For the modeling and fitting of the sources, we use the HAWC accelerated likelihood (HAL)~\footnote{HAL; \url{https://github.com/threeML/hawc_hal}} plugin with the Multi-Mission Maximum Likelihood (3ML)~\footnote{3ML; \url{https://github.com/threeML/threeML}} framework \citep{HAL,threeML}. For the source analysis, we first select a disk ROI centred at ($l=40.42^{\circ}$, $b=-0.85^{\circ}$) with a radius of $2.5^{\circ}$. We then determine the number of gamma-ray sources within our ROI. Finally, we find for each of these sources the best spectral and morphological models by comparing a statistical measure, named test statistic (TS), defined as:

\begin{equation}
   \mathrm{TS} = 2\,\mathrm{ln}\left(\frac{L_{1}}{L_{0}}\right).
\end{equation}
In this equation, TS compares the likelihood of an alternative hypothesis ($L_{1}$) to that of a null (background-only) hypothesis ($L_{0}$).
Analyses done with HAWC data often use TS to determine a pre-trial statistical significance of a fitted model inside a chosen ROI. The model may contain any number of free parameters depending on the number of sources and their spectral and morphological shapes being fitted. A pre-trial statistical significance measured in Gaussian standard deviation units can be estimated based on the Wilks' theorem \citep{Wilks}, assuming one degree of freedom:

\begin{equation}\label{eqn:sigma}
   \sigma \simeq \sqrt{\mathrm{TS}}.
\end{equation}
Note that the Wilks' Theorem is valid for HAWC data \citep{2HWC}.

\subsection{Source Search Pipeline} \label{subsec:systematicsourcesearch}

In \citet{Abeysekara:2018qtj}, a point source model was fitted at each of the known X-ray emission regions ``e1'' and ``w1'' \citep{Safi-Harb_1997} for the east and west jet lobes, respectively, along with an extended source model for MGRO~J1908+06 at its known location. Then, a residual map was examined to confirm that the model explained the excess emission over the background.  
With the increased statistics in the data used for this work, we no longer use prior information about the source location or the number of sources in the region. Instead, we carry out a source search pipeline within our chosen ROI to identify the model that best describes the observational data. 

We first add point sources iteratively to a baseline model until the most significant excess emission in our residual map becomes less than ${\rm TS}=16$ \cite{4HWCCatalog}. Then, each of the point sources found in the previous step is iteratively replaced by an extended source. Each newly found model is compared with the current best model. A model that improves the fit by $\Delta {\rm TS} > 16$ is accepted as the new best model. During a successful extension test, if any of the remaining point sources become insignificant (TS $<16$), they are removed from the new best model.

Our baseline model includes a diffuse background emission (DBE) ``source'' fitted between $l=38^{\circ}$ and $l=43^{\circ}$. This extended source is designed to account for unresolved sources and the Galactic diffuse emission. The DBE is described as a source with a one dimensional Gaussian profile along the Galactic equator.

In the source search pipeline, the spectral model is assumed to be a simple power law, 
\begin{equation}\label{eq:powerlaw}
    \frac{\mathrm{d}N}{\mathrm{d}E_{\gamma}} = K\left(\frac{E_{\gamma}}{E_{\mathrm{piv}}}\right)^{\alpha},
\end{equation} 
where $K$ is flux normalization at pivot energy $E_{\mathrm{piv}}$ and spectral index $\alpha$. For the pivot energy, we have used $E_{\rm piv}=10$~TeV which has minimum correlation between free parameters.
When an extended source is added, a Gaussian template is used to describe the spatial morphology (see Appendix~\ref{app:sec:morphs}).

\begin{table*}[ht]
\centering
\footnotesize
\begin{tabular}{c|c|c|c|c}
\multicolumn{1}{c|}{\bf Source} &
\multicolumn{1}{c|}{\bf RA, Dec [$^{\circ}$, $^{\circ}$]} &
\multicolumn{1}{c|}{\bf K [TeV$^{-1}$cm$^{-2}$s$^{-1}$]} &
\multicolumn{1}{c|}{\bf $\alpha$} &
\multicolumn{1}{c}{\bf TS} \\
\hline
&&&\\
West lobe & $287.61^{+0.02}_{-0.02}$, $5.06^{+0.02}_{-0.02}$ & $2.02^{+0.32}_{-0.31}\times10^{-15}$ & $-2.35^{+0.12}_{-0.11}$ & $79.5$ \\
&&&\\
East lobe & $288.41^{+0.02}_{-0.02}$, $4.93^{+0.02}_{-0.02}$ & $1.65^{+0.28}_{-0.28}\times10^{-15}$ & $-2.44^{+0.13}_{-0.12}$ & $53.9$ \\
\end{tabular}
\caption{\label{table:results} Best fit results of the SS~433 lobes using point source models and simple power law spectral models (Equation~\ref{eq:powerlaw}). $E_{\textrm{piv}}=10$~TeV is used for the fit. Only the statistical uncertainties are indicated in this table.}
\end{table*}

\begin{table*}
\centering
\begin{tabular}{c c c c c c c c}
\multicolumn{1}{l}{\bf Source} &
\multicolumn{1}{c}{\bf RA, Dec} &
\multicolumn{1}{c}{\bf $K$} &
\multicolumn{1}{c}{\bf $\alpha$} &
\multicolumn{1}{c}{\bf $\beta$} &
\multicolumn{1}{c}{\bf Extension} &
\multicolumn{1}{c}{\bf TS} &
\multicolumn{1}{c}{\bf 1LHAASO}
\\
\multicolumn{1}{l}{\bf (HAWC)} &
\multicolumn{1}{c}{\bf [$^{\circ}$, $^{\circ}$]} &
\multicolumn{1}{c}{\bf [TeV$^{-1}$cm$^{-2}$s$^{-1}$]} &
&
&
\multicolumn{1}{c}{\bf [$^\circ$]} &
&
\multicolumn{1}{c}{\bf Counterpart}
\\
\hline\hline
J1902+0656${}^{*}$ & $285.54^{+0.03}_{-0.03}$, $6.94^{+0.04}_{-0.03}$ & $9.61^{+2.62}_{-2.53}\times10^{-16}$ & $-2.49^{+0.20}_{-0.19}$ & - & - & 20.0 & J1902+0648 \\
J1905+0711${}^{*}$ & $286.46^{+0.04}_{-0.04}$, $7.18^{+0.04}_{-0.05}$ & $8.95^{+2.76}_{-2.66}\times10^{-16}$ & $-2.73^{+0.28}_{-0.27}$ & - & - & 12.7 & J1906+0712 \\
J1908+0552 & $286.91^{+0.03}_{-0.03}$, $5.87^{+0.03}_{-0.03}$ & $1.44^{+0.35}_{-0.33}\times10^{-15}$ & $-2.19^{+0.15}_{-0.15}$ & - & - & 30.3 & \nodata \\
J1908+0618 & $287.04^{+0.02}_{-0.02}$, $6.30^{+0.01}_{-0.01}$ & $9.04^{+0.48}_{-0.45}\times10^{-14}$ & $-2.42^{+0.02}_{-0.02}$ & $0.18^{+0.02}_{-0.01}$ & $1.54^{+0.07}_{-0.07}$ & 2398.9 & J1908+0621 \\
\hline
\end{tabular}
\caption{Best fit results of the background sources. 1LHAASO counterparts based on their positions are indicated. Note only the statistical uncertainties are indicated here. $\beta$ is the parameter of log parabola spectral model described in Appendix~\ref{app:sec:spectra}. Note * indicates source candidates.}
\label{table:bkgresults}
\end{table*}

\section{Results} \label{sec:results}

\subsection{Significance and Morphology}
Figure~\ref{fig:sigmap}~(a) shows the map of the SS~433 region in Galactic coordinates, produced with 2,565~days of HAWC data. The color bar indicates the pre-trial significance defined in Equation~\ref{eqn:sigma}. Known TeV gamma-ray sources in the region are labeled \citep{TeVCat}.

Our source search pipeline (as described in Section~\ref{subsec:systematicsourcesearch}) finds six sources. They include one extended source HAWC~J1908+0618, which corresponds to MGRO~J1908+06, one point source that coincides with the east jet lobe of SS~433 and another point source that matches with the known location of the west jet lobe of SS~433. We have also found three additional point sources compared to \citet{Abeysekara:2018qtj}, namely, HAWC~J1902+0656, HAWC~J1905+0711, and HAWC~J1908+0552, resulting from the increased amount of data and the use of a full disk ROI compared to the semi-disk ROI \citep{Abeysekara:2018qtj}. 

We test three different morphological models for HAWC~J1908+0618, Gaussian, electron diffusion~\citep{HAWCGeminga}, and inverse power law. Appendix~\ref{app:sec:morphs} has the details on the morphological models. 

Table~\ref{table:results} shows the best-fit position, flux normalization, spectral index, and TS values while assuming the simple power law spectral model (Equation~\ref{eq:powerlaw}) for each SS~433 jet lobe. Table~\ref{table:bkgresults} shows the best-fit parameters for the four background sources. Among these sources, HAWC~J1902+0656 and HAWC~J1905+0711 are reported in this work as source candidates since they have TS $<25$, but they have good spatial agreement with sources from the first Large High Altitude Air Shower Observatory catalog (1LHAASO) \citep{1LHAASOCat}. HAWC~J1905+0711 has TS below 16 because while it was identified as a source in the source search pipeline, adopting an electron diffusion morphology for HAWC~J1908+0618 after the source search process has decreased its statistical significance. 

Figure~\ref{fig:sigmap}~(b) shows the significance map zoomed in around the known location of the central binary, SS~433. Note that the best-fit emission from background sources from Table~\ref{table:bkgresults} have been subtracted. The green contour lines are the X-ray observations from R{\"o}ntgensatellit (ROSAT; \citealp{1996A&A...312..306B}), showing emissions from three distinctive regions, namely, the central binary, the west jet lobe (closer to the Galactic equator), and the east jet lobe.
The crosses indicate some of the key interaction regions observed in 2-10~keV X-rays \citep{Safi-Harb_1997}, including ``w1'' and ``w2'' for the west jet lobe and ``e1'', ``e2'', and ``e3'' for the east jet lobe. Figure~\ref{fig:sigmap}~(b) shows that the TeV gamma-ray hotspots are well-aligned with the X-ray jets. 

The red cross at the center is the known location of the binary system. No significant emission from the central binary is detected by HAWC. The 95\% confidence level upper limit on the flux of the central binary measured at 10~TeV is $1.38\times10^{-15}~\mathrm{TeV^{-1}cm^{-2}s^{-1}}$ for a spectral index of $-2.5$.

\begin{figure*}[ht!]
\centering
\gridline{\fig{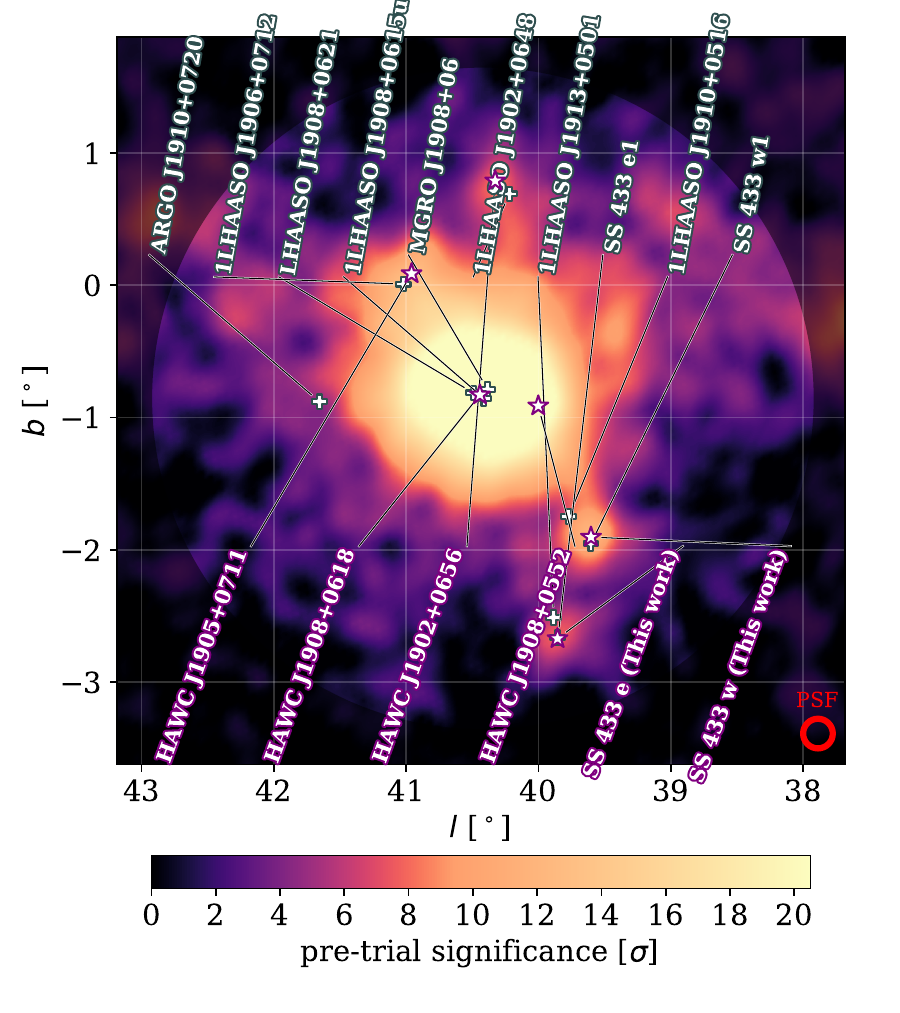}{0.66\columnwidth}{(a)} \fig{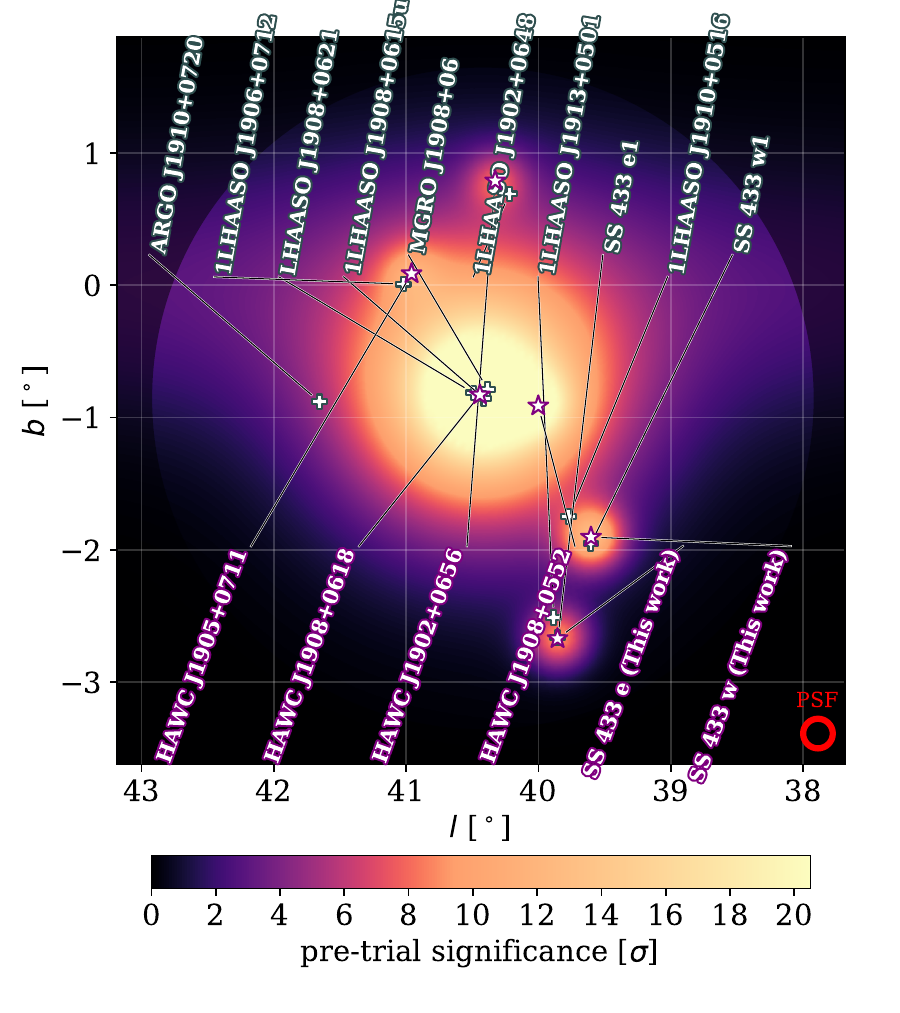}{0.66\columnwidth}{(b)} \fig{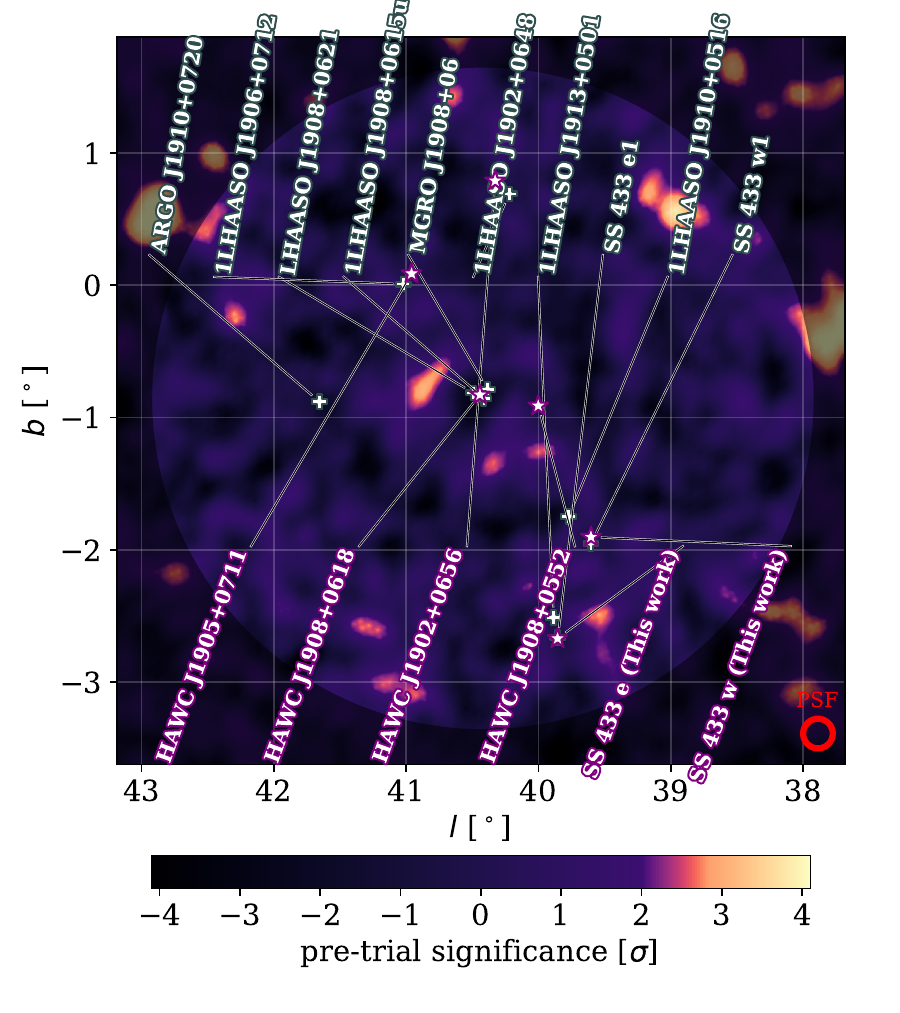}{0.66\columnwidth}{(c)}}
\caption{Significance maps of the results from the source search pipeline. The lower labels indicate the sources found by this analysis. (a)HAWC data map. (b) Model map. (c) Residual map. 
\label{fig:sigmap_modeling}}
\end{figure*}

Figure~\ref{fig:sigmap_modeling}~(b) shows the nested model map that contains the six sources and DBE that we have obtained as the best model for our ROI. 
Figure~\ref{fig:sigmap_modeling}~(c) presents the residual map that has the final model subtracted from the HAWC significance map (Figure~\ref{fig:sigmap_modeling}~(a)). The residual map has no obvious excess left in the ROI as expected, which is confirmed separately using a one-dimensional significance histogram.

\begin{figure}[ht!]
    \centering
    \includegraphics[width=\columnwidth]{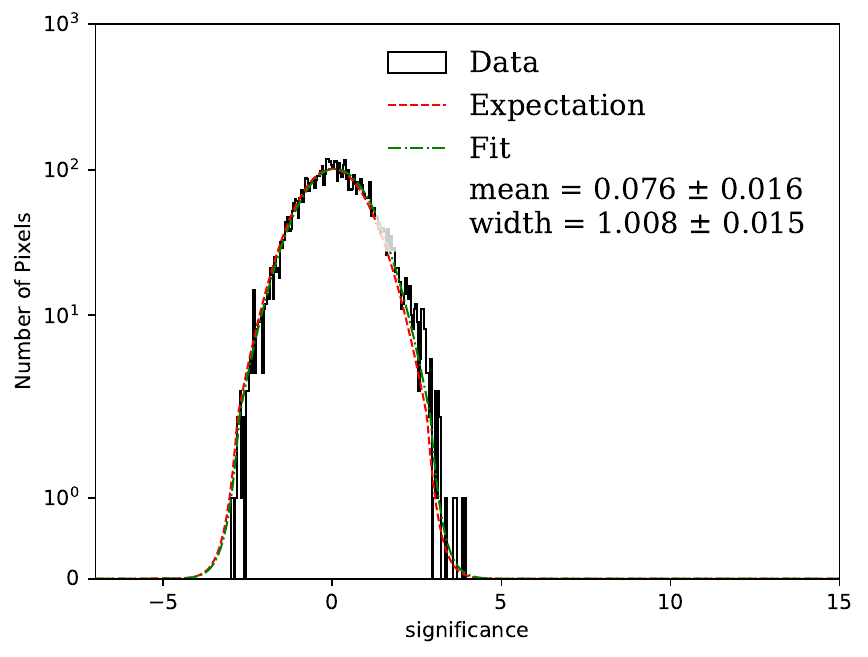}
    \caption{A one-dimensional significance histogram of the final residual map shown in Figure~\ref{fig:sigmap_modeling}~(c).  
    \label{fig:sighist}}
\end{figure}

Figure~\ref{fig:sighist} shows the one-dimensional significance histogram of the residual map by collecting the significance values associated with each of the pixels within the ROI. With all the excess fitted and subtracted, the distribution should resemble a background-only distribution. The expected background-only distribution is indicated with a red dashed curve, which is a Gaussian centered at 0 between $-5\sigma$ and $5\sigma$.

\subsection{Spectral Studies} \label{subsec:spectralstudies}

The newly produced HAWC dataset with improved algorithms and more statistics allows us to study the spectrum of each lobe in detail. We divide the data into 10 NN energy bins from ``c'' to ``l'' \citep{Abeysekara:2019edl}. The true photon energy range per energy bin depends on the fitted source itself. 
 
We fit a simple power-law (Equation~\ref{eq:powerlaw}) to the data corresponding to each energy bin while fixing the spectral index to the best-fit values from Table~\ref{table:results}, $\alpha = - 2.35$ and $\alpha = - 2.44$ for the east and west lobes, respectively. Further information on the energy bins used in this work can be found in Appendix~\ref{app:sec:energy_bins}.

Figure~\ref{fig:hawc_spectra} presents the best fit spectrum of each lobe. As shown in the figure, bins ``f'', ``g'', ``h'', and ``i'' are significant enough to calculate flux points for the east jet lobe, while upper limits are computed from the rest of the bins. The dotted line indicates a simple power law that is fitted to the entire energy range. As for the west jet lobe, bins ``f'', ``g'', ``h'', ``i'', and ``k'' have flux points. While the fitted spectrum of the east lobe is consistent with the flux point from \citet{Abeysekara:2018qtj}, the fitted spectrum of the west lobe at 20~TeV is $3\sigma$ away from the previous work. Finally, the fitted spectra of the two lobes from this work are consistent with each other.

Applying alternative spectral models for the lobes, including the cutoff power-law (COPL) and log parabola (LP) models, do not show significant improvements to the fit, hence we conclude that the power law best describes the spectra of the two jet lobes. $\Delta {\rm TS_{COPL}}=8$ and $\Delta {\rm TS_{LP}}=9$, respectively, for the east jet lobe and $\Delta {\rm TS_{COPL}}=7$ and $\Delta {\rm TS_{LP}}=6$, respectively, for the west jet lobe. The description of COPL and LP can be found in Appendix~\ref{app:sec:spectra}.

\begin{figure*}[ht!]
\centering
\gridline{\fig{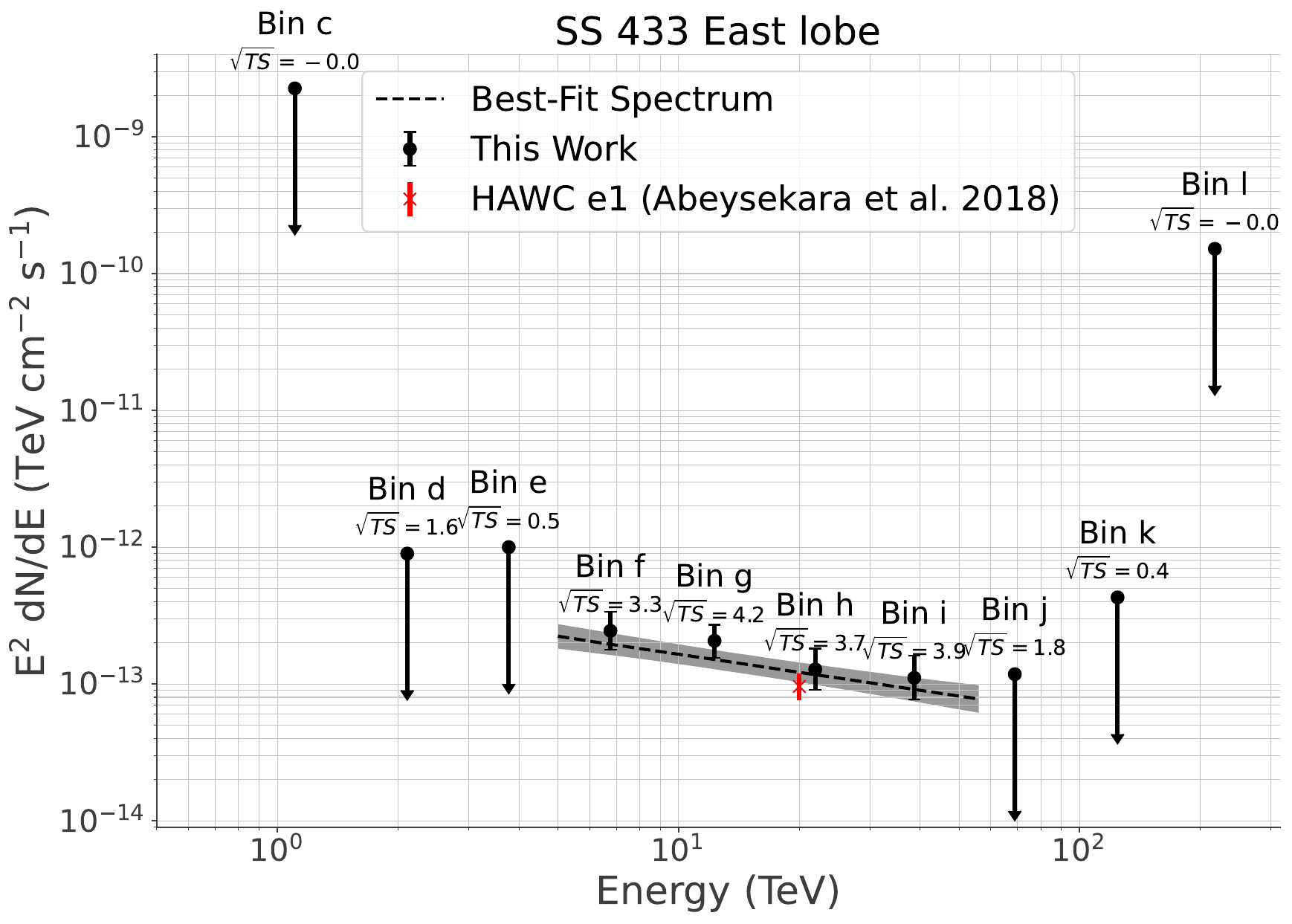}{\columnwidth}{(a)} \fig{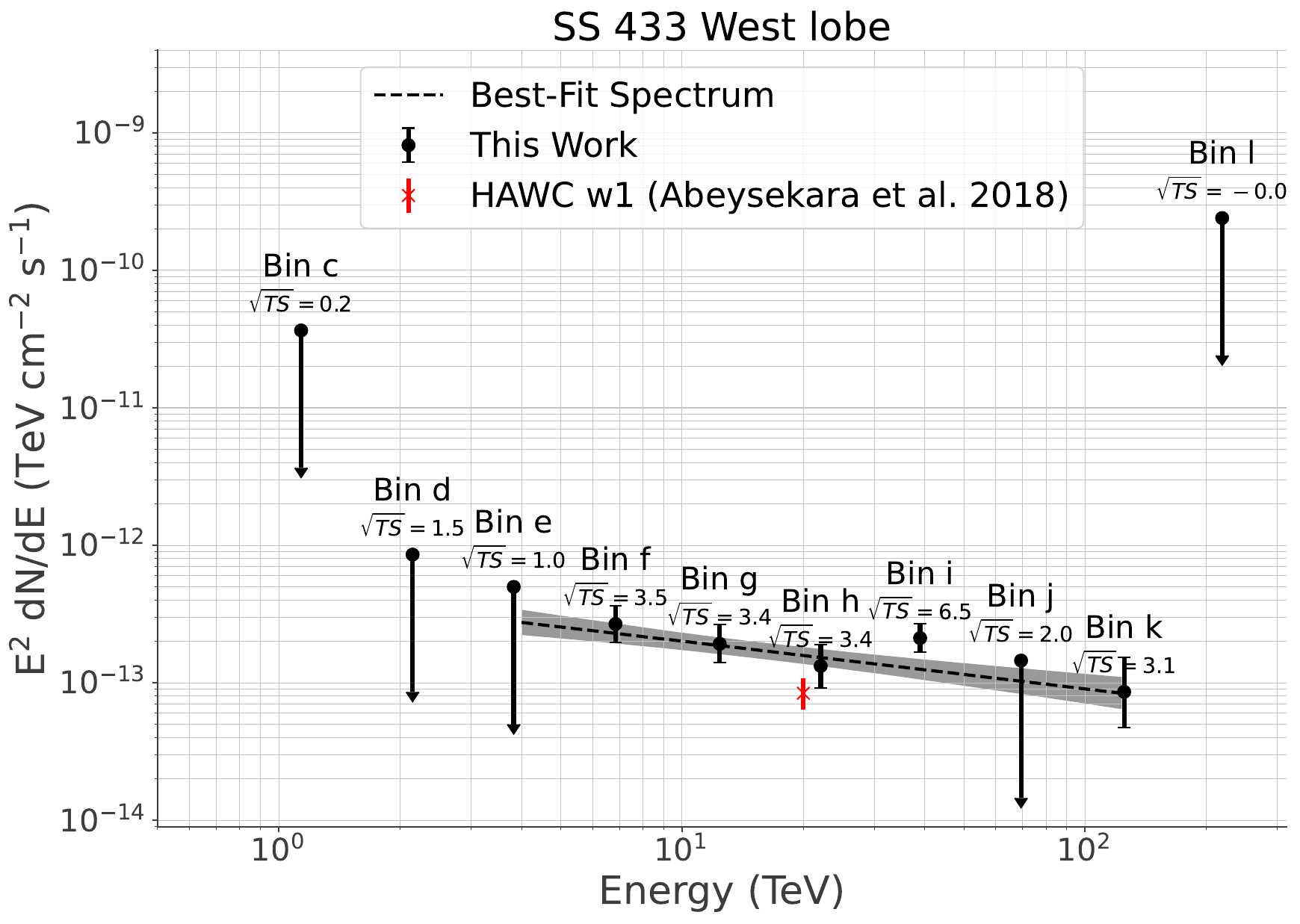}{\columnwidth}{(b)}}
\caption{The best-fit power-law spectra of the SS~433 lobes. Information on each data point corresponding to energy bins is presented in Appendix~\ref{app:sec:energy_bins}. The shaded region represents the $1\,\sigma$ statistical error band. The red data point at 20~TeV represents the best-fit result at e1 (w1) for the east (west) lobe from \citet{Abeysekara:2018qtj}.
\label{fig:hawc_spectra}}
\end{figure*}

\subsection{Extension and Energy Range of Jet Lobes}

In this work, the SS~433 east and west jet lobes are best described using two point source models. We use the $5\,\sigma$ contours to constrain the regions that produce significant VHE gamma-ray emission. Figure~\ref{fig:sigmap}~(b) shows the $5\,\sigma$ contours in white. When compared with the X-ray contours in green, we can see that the HAWC observations are constrained to the ``head'' area closer to the central binary system for both of the lobes. In terms of the X-ray emission regions, both ``e1'' and ``w1'' are located close to the best-fit positions and the centers of the contours. Therefore, we confirm that the use of the known X-ray emission regions in \citet{Abeysekara:2018qtj} is justified.

Furthermore, we study the energy range of the gamma-ray photons observed by HAWC. The maximum photon energy at the $1\,\sigma$-level for the east lobe is 56~TeV, which is significantly higher than the maximum photon energy of 25~TeV reported in ~\citet{Abeysekara:2018qtj}. The minimum photon energy at the $1\,\sigma$-level is 5~TeV. As for the west lobe, we compute the maximum photon energy to be 123~TeV and the minimum energy as 4~TeV.

\subsection{Energy Dependent Morphologies of Jet Lobes}
\label{sec:energymorph}

\begin{figure*}[ht!]
\centering
\gridline{\fig{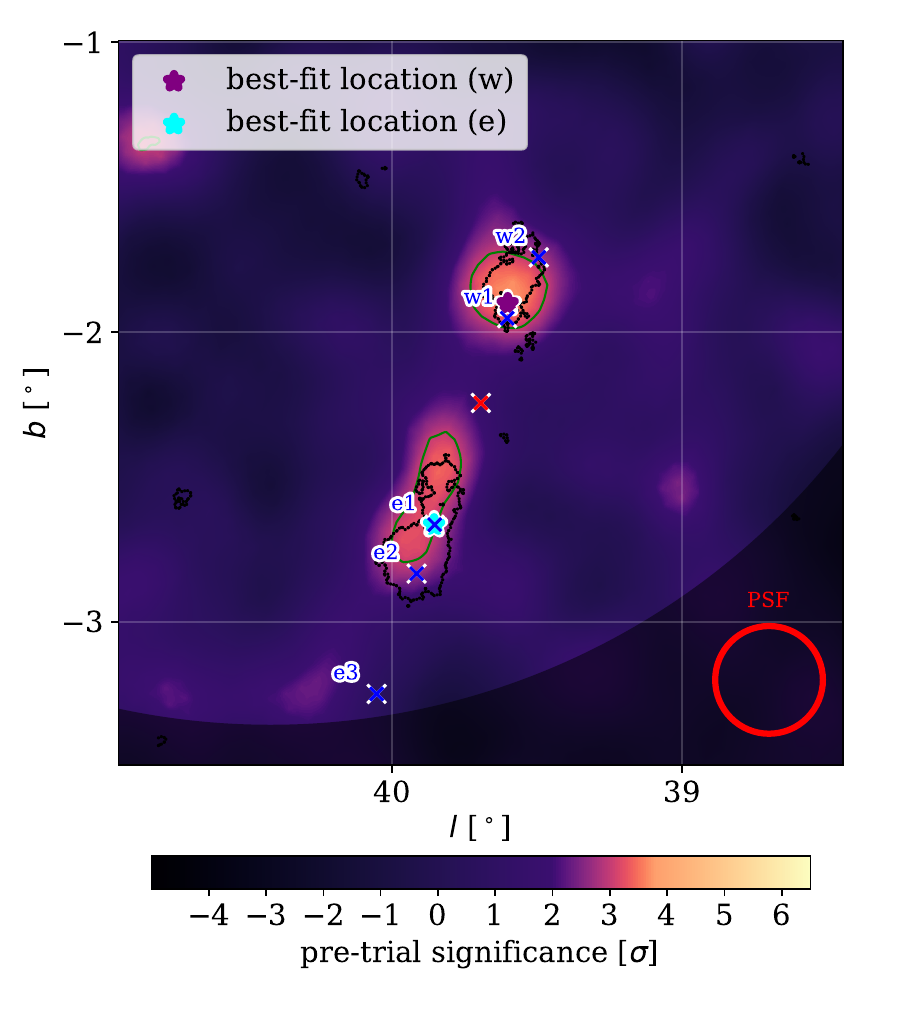}{0.66\columnwidth}{(a) $1-10~\mathrm{TeV}$} \fig{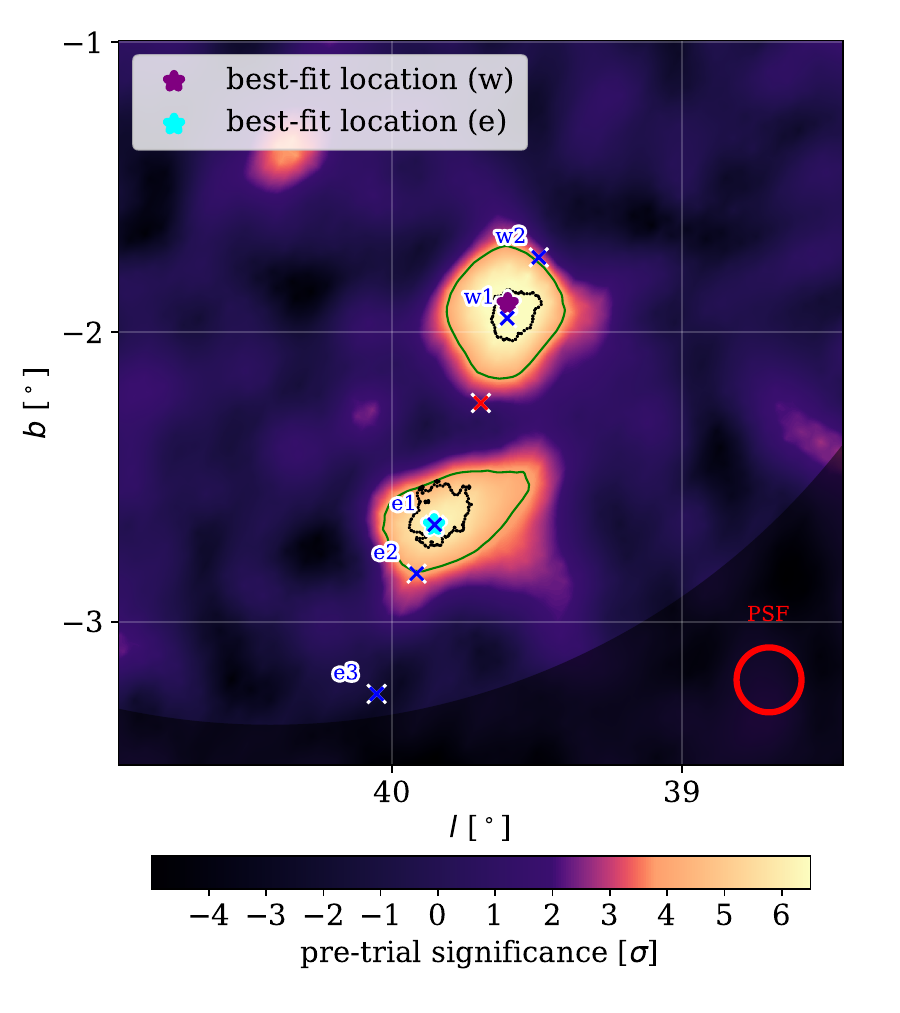}{0.66\columnwidth}{(b) $>10~\mathrm{TeV}$} \fig{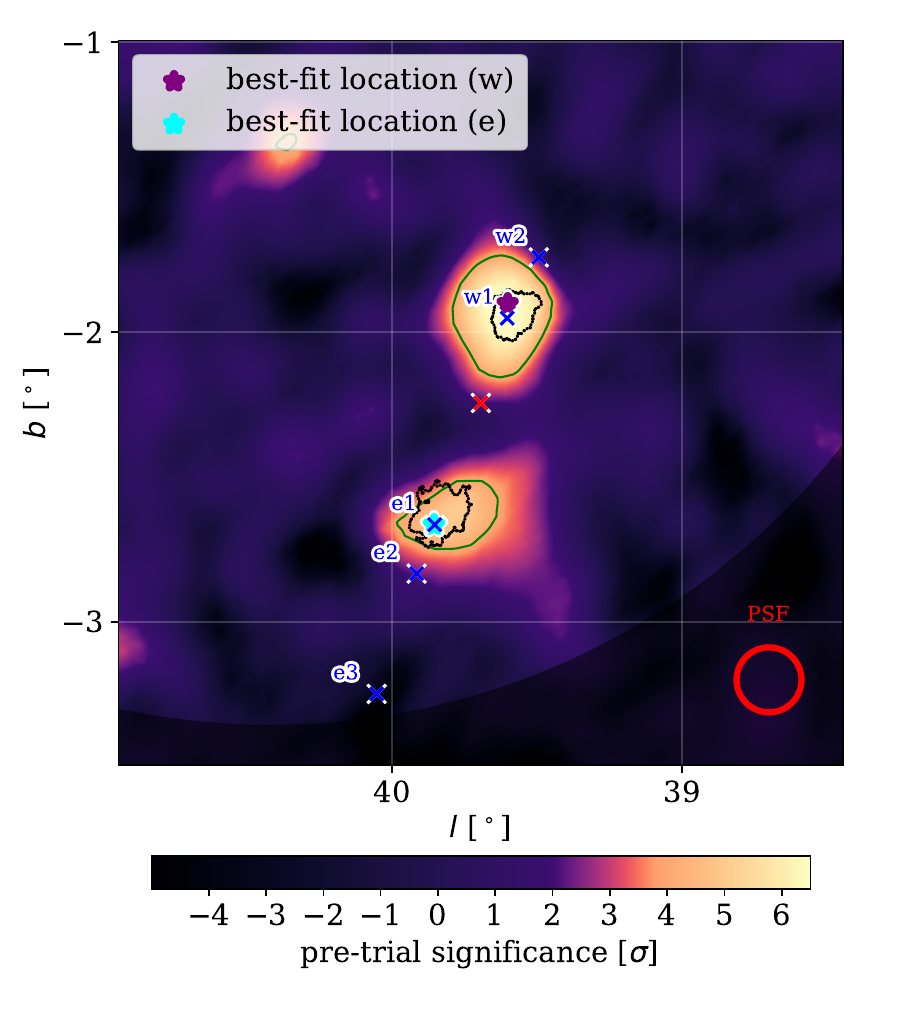}{0.66\columnwidth}{(c) $>18~\mathrm{TeV}$}}
\caption{Significance maps around the SS~433 jet lobes for different energy ranges. The background sources have been subtracted.  
The red circle in each panel indicates the angular resolution of the bin that includes the most number of events for the given energy range.
(a) shows the significance map corresponding to $1-10$~TeV. The green contour indicates the $3\sigma$ level in the HAWC data. The black contour indicates the $3\sigma$ level in the H.E.S.S. data at energies $2.5-10~\mathrm{TeV}$. (b) shows the significance map corresponding to above $10$~TeV. The green contour indicates the $4\sigma$ level in the HAWC data at energies above $10~\mathrm{TeV}$. The black contour indicates the $4\sigma$ level in the H.E.S.S. data at energies above $10~\mathrm{TeV}$. (c) shows the significance map corresponding to above $18~\mathrm{TeV}$. The green contour indicates the $4\sigma$ level in the HAWC data. The black contour indicates the $4\sigma$ level in the H.E.S.S. data at energies above $10~\mathrm{TeV}$.  \citep{HESS24} 
\label{fig:ss433_edep_sigmaps}}
\end{figure*}

\movetableright=-2cm
\begin{table*}[ht!]
\footnotesize
\centering
\begin{tabular}{c|ccc|ccc}
\multicolumn{1}{c|}{} &
\multicolumn{3}{c|}{West Lobe} &
\multicolumn{3}{c}{East Lobe} \\
\multicolumn{1}{c|}{Energy} &
\multicolumn{1}{c}{\bf RA} &
\multicolumn{1}{c}{\bf Dec} &
\multicolumn{1}{c|}{\bf TS} &
\multicolumn{1}{c}{\bf RA} &
\multicolumn{1}{c}{\bf Dec} & 
\multicolumn{1}{c}{\bf TS}  \\
\hline
\multicolumn{1}{c|}{$1 - 10~\mathrm{TeV}$} & $287.57\pm0.05$ & $5.06\pm0.05$ & $15$ & $288.42\pm0.05$ & $4.98\pm0.05$ & $13$ \\
\multicolumn{1}{c|}{$>10~\mathrm{TeV}$} & $287.61\pm0.02$ & $5.06\pm0.02$ & $70$ & $288.41\pm0.02$ & $4.93\pm0.02$ & $48$ \\
\multicolumn{1}{c|}{$>18~\mathrm{TeV}$} & $287.65\pm0.02$ & $5.06\pm0.02$ & $60$ & $288.40\pm0.03$ & $4.92\pm0.03$ & $31$
\end{tabular}
\caption{\label{table:energy_dependent_positions} For each lobe, best-fit positions and TS values are presented for the energy ranges presented in Figure~\ref{fig:ss433_edep_sigmaps}.}
\end{table*}

H.E.S.S. observed energy dependent morphological changes to the SS~433 jet lobes \citep{HESS24}. We have divided our data into three energy bins to look for similar changes. Figure~\ref{fig:ss433_edep_sigmaps} features the significance maps corresponding to: (a) $1<E<10$~TeV; (b) $E>10$~TeV; and (c) $E>18$~TeV. The maps contain contours from the H.E.S.S. observations \citep{HESS24}, which are consistent with the emission observed by HAWC. While the best-fit position of the east lobe is located nearer to the central binary at higher energies, the change is smaller than the angular resolution of HAWC. We note that the emission at higher energies are clearly centered around e1 and w1 regions. Furthermore, Table~\ref{table:energy_dependent_positions} provides the best fit positions of the lobes for each energy range. The best-fit positions computed for the lobes are consistent between the energy bins.

\subsection{Systematic Studies}
\label{sec:systematics}

The systematic uncertainties associated with our analysis are summarized in Table~\ref{table:systematics}.
The systematic uncertainties contain the effects from the detector simulations \citep[see][]{2017ApJ...843...39A, Abeysekara:2019edl}. The most dominant effect is the late light component, which indicates uncertainties from the discrepancy between the arrival times of light from an actual air shower and the arrival times of light from the laser pulse used for calibrations. The sources of systematic uncertainties from Table~\ref{table:systematics} are described in \citet{Abeysekara:2019edl} in more detail.

\movetableright=-2cm
\begin{table*}[ht!]
\footnotesize
\centering
\begin{tabular}{c|cc|cc|cc}
\multicolumn{1}{c|}{\bf Type} &
\multicolumn{2}{c|}{\bf RA, Dec} &
\multicolumn{2}{c|}{\bf Flux Normalization} &
\multicolumn{2}{c}{\bf $\alpha$} \\
\multicolumn{1}{c|}{} & 
\multicolumn{2}{c|}{[${}^{\circ}$, ${}^{\circ}$]} &
\multicolumn{2}{c|}{\bf[$10^{-16}$ TeV${}^{-1}$ cm$^{-2}$ s$^{-1}$]} &
\multicolumn{2}{c}{} \\
\multicolumn{1}{c|}{} &
\multicolumn{1}{c}{\bf west} &
\multicolumn{1}{c|}{\bf east} &
\multicolumn{1}{c}{\bf west} &
\multicolumn{1}{c|}{\bf east} &
\multicolumn{1}{c}{\bf west} &
\multicolumn{1}{c}{\bf east} \\
\hline
\multicolumn{1}{c|}{Late light} & $\pm0.001, \pm0.0007$ & $\pm0.0006, \pm0.0008$ & $\pm1.84$ & $\pm1.55$ & $\pm0.01$ & $\pm0.02$ \\
\multicolumn{1}{c|}{Charge uncertainty} & $\pm0.001, \pm0.002$ & $\pm0.0003, \pm0.002$ & $\pm1.06$ & $\pm0.73$ & $\pm0.01$ & $\pm0.02$ \\
\multicolumn{1}{c|}{PMT threshold} & $\pm0.0002, \pm0.0002$ & $\pm0.0006, \pm0.0002$ & $\pm0.53$ & $\pm0.34$ & $\pm0.005$ & $\pm0.006$ \\
\multicolumn{1}{c|}{PMT efficiency} & $\pm0.0005, \pm0.002$ & $\pm0.0008, \pm0.001$ & $\pm0.35$ & $\pm0.50$ & $\pm0.02$ & $\pm0.02$ \\
\hline
\multicolumn{1}{c|}{\bf Quadratic Sum} & $\pm0.002, \pm0.003$ & $\pm0.001, \pm0.002$ & $\pm2.22$ & $\pm1.82$ & $\pm0.03$ & $\pm0.04$
\end{tabular}
\caption{\label{table:systematics} Detector systematic effects measured for four different categories and their quadratic sums.}
\end{table*}

\section{Discussion and Conclusion} \label{sec:conclusion}

\begin{figure}[ht!]
\centering
    \includegraphics[width=0.99\columnwidth]{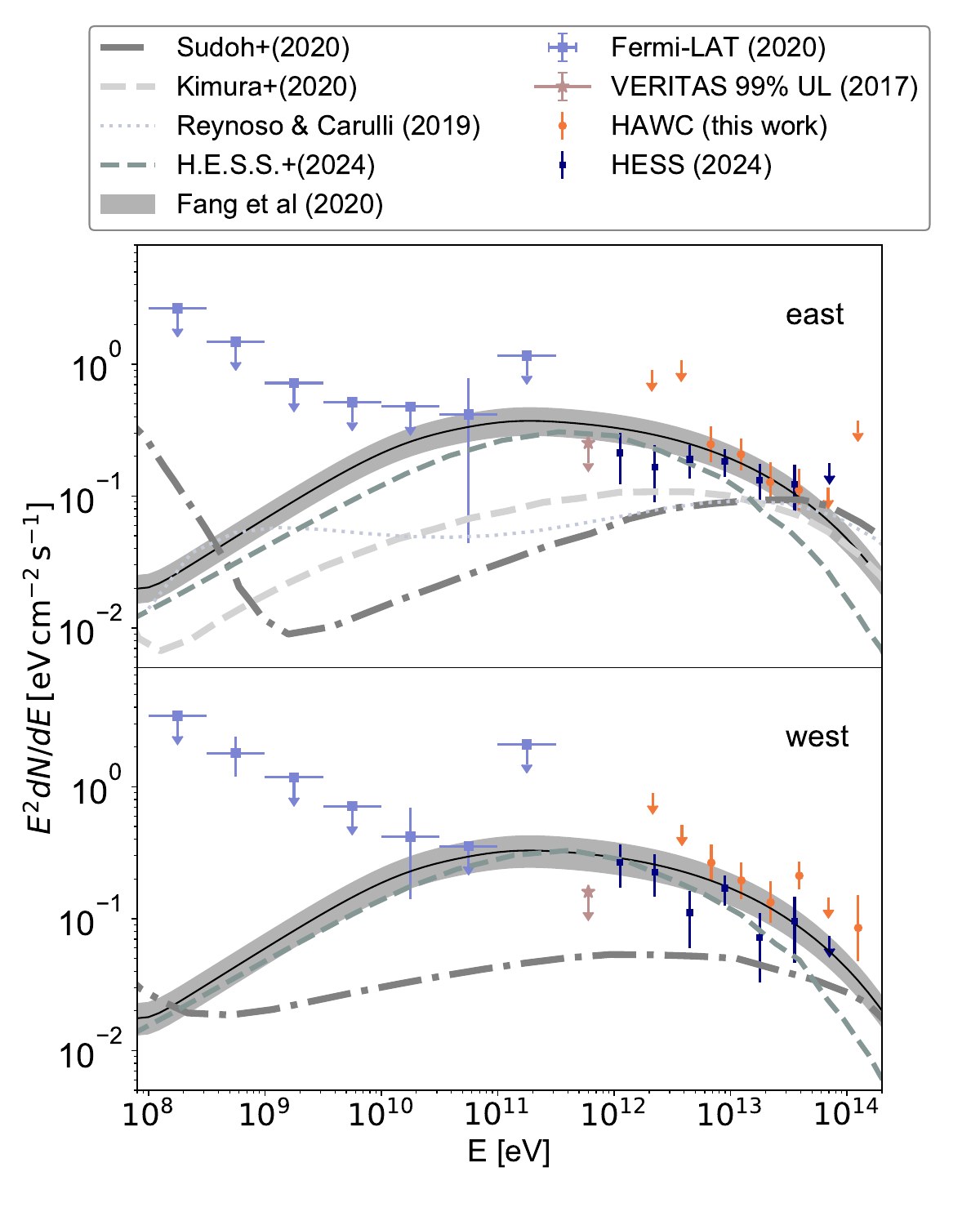}
    \caption{The spectral energy distribution of the gamma-ray emission site in the SS 433 east (top) and west (bottom) lobes measured by HAWC (this work; orange), {\it Fermi}-LAT (\citealp{Fang_2020}, light blue markers), H.E.S.S. (\citealp{HESS24}, navy square markers), and VERITAS (\citealp{Kar:2017oe}, brown upper limit). Models are shown for reference, including \citet{Reynoso:2019vrp} (parameter 3a; light grey dotted curve), \citet{Sudoh:2019jup} (dark grey dash-dotted curves),  \citet{Fang_2020} (grey shaded regions), \citet{Kimura:2020acy} (scenario D; light grey dashed curve), and \citet{HESS24} (grey dashed curve). \label{fig:sed}}
\end{figure}

We have analyzed 2,565~days of HAWC data to study the TeV gamma-ray emission from the jet lobes of SS~433. The new data contain more and improved statistics to carry out a thorough blind search of the region. Our blind search successfully identifies two point sources located close to the X-ray hotspots, ``e1'' and ``w1'', which are known as jet interaction regions for the east and west jet lobes, respectively. However, we find no evidence of extended emission for the lobes. The blind search within the ROI has also identified four background sources of which HAWC~J1908+0618 is the only extended source that is best described by the electron diffusion morphological model, similar to \citet{Abeysekara:2018qtj}.

Our spectral studies find that the sources can be described by a power law spectrum with no cut off. The $1\sigma$ bound of the maximum photon energy has increased from 25~TeV to 56~TeV and 123~TeV for the east and west jet lobes, respectively, compared to our previous analysis \citep{Abeysekara:2018qtj}. The spectra are measured to be softer than previously assumed in \citet{Abeysekara:2018qtj}.

When fitting to the broadband energy spectrum using an electron spectrum that follows a power-law with an exponential cutoff as in \citet{Abeysekara:2018qtj}, we find the total energy of injected electrons as $W_e = 1.3\times 10^{47}\,\rm erg$, well within the energy budget of the kinetic power of the jets \citep{2004ASPRv..12....1F}.

The spectra in Figure~\ref{fig:hawc_spectra} constructed from fitting the data that are divided into NN energy bins are consistent with the flux points at 20~TeV from \citet{Abeysekara:2018qtj}. The best-fit spectral indices $\alpha=-2.44^{+0.13}_{-0.12}$ and $\alpha=-2.35^{+0.12}_{-0.11}$ are lower than what we assumed in \citet{Abeysekara:2018qtj} for which the spectral indices were fixed to $\alpha=-2.0$. The pivot energy used in this work, $E_{\mathrm{piv}}=10$~TeV, is also lower than $E_{\mathrm{piv}}=20$~TeV used in \citet{Abeysekara:2018qtj}.

Figure~\ref{fig:sed} presents the spectral energy distribution of the gamma-ray emission from the eastern and western SS~433 jet lobes between 0.1~GeV and 100~TeV. Models have been proposed to explain the TeV emission by SS~433, invoking continuous or instantaneous injections of electrons and protons from the synchrotron knots \citep{Reynoso:2019vrp, Sudoh:2019jup, Fang_2020}, as well as a standing shock in the outer jets \citep{HESS24}. When comparing our observation with these recent models, we find that the models confined by previous IACT limits are generally too low to explain the HAWC and H.E.S.S. observations.   The grey shaded region in the figure corresponds to the best-fit gamma-ray model in \citet{Fang_2020} obtained by fitting a steady electron injection model as in \citet{Abeysekara:2018qtj} to the {\it Fermi}-LAT and HAWC data from \citet{Abeysekara:2018qtj}.
The grey dashed curve is a model from \citet{HESS24} obtained by fitting a similar electron injection model to the H.E.S.S. and X-ray data. The model is in tension with the HAWC measurements above $\sim$30~TeV, suggesting that the multi-wavelength production mechanism may be more complicated than a one-zone model where electrons simultaneously produce X-ray and TeV gamma-ray emission.

The successful identification of SS~433 through a blind search with the HAWC data confirms the existence of gamma-ray emission by the microquasar and pinpoints the gamma-ray production sites inside the large-scale jet. The photon energy reaches beyond 100~TeV, suggesting that electrons must have been efficiently accelerated given the short cooling time at such extreme energies \citep{Abeysekara:2018qtj, Fang_2020}. More discoveries of very-high-energy gamma-ray-emitting jets by future air shower gamma-ray detectors will help understand the mechanism of particle acceleration at large distances from the compact object.

\section{Acknowledgements}
We acknowledge the support from: the US National Science Foundation (NSF); the US Department of Energy Office of High-Energy Physics; the Laboratory Directed Research and Development (LDRD) program of Los Alamos National Laboratory; Consejo Nacional de Ciencia y Tecnolog\'{i}a (CONACyT), M\'{e}xico, grants 271051, 232656, 260378, 179588, 254964, 258865, 243290, 132197, A1-S-46288, A1-S-22784, CF-2023-I-645, c\'{a}tedras 873, 1563, 341, 323, Red HAWC, M\'{e}xico; DGAPA-UNAM grants IG101323, IN111716-3, IN111419, IA102019, IN106521, IN114924, IN110521 , IN102223; VIEP-BUAP; PIFI 2012, 2013, PROFOCIE 2014, 2015; the University of Wisconsin Alumni Research Foundation; the Institute of Geophysics, Planetary Physics, and Signatures at Los Alamos National Laboratory; Polish Science Centre grant, DEC-2017/27/B/ST9/02272; Coordinaci\'{o}n de la Investigaci\'{o}n Cient\'{i}fica de la Universidad Michoacana; Royal Society - Newton Advanced Fellowship 180385; Generalitat Valenciana, grant CIDEGENT/2018/034; The Program Management Unit for Human Resources \& Institutional Development, Research and Innovation, NXPO (grant number B16F630069); Coordinaci\'{o}n General Acad\'{e}mica e Innovaci\'{o}n (CGAI-UdeG), PRODEP-SEP UDG-CA-499; Institute of Cosmic Ray Research (ICRR), University of Tokyo; National Research Foundation of Korea (RS-2023-00280210, NRF-2021R1A2C1094974). H.F. acknowledges support by NASA under award number 80GSFC21M0002. We also acknowledge the significant contributions over many years of Stefan Westerhoff, Gaurang Yodh and Arnulfo Zepeda Dom\'inguez, all deceased members of the HAWC collaboration. Thanks to Scott Delay, Luciano D\'{i}az and Eduardo Murrieta for technical support.

C.D.~Rho analyzed the data and prepared the original manuscript. Y.~Son analyzed the data and performed the maximum likelihood analysis. K.~Fang carried out the interpretation of results and the discussion section calculations. The full HAWC collaboration has contributed through the construction, calibration, and operation of the detector, the development and maintenance of reconstruction and analysis software, and vetting of the analysis presented in this manuscript. All authors have reviewed, discussed, and commented on the results and the manuscript.

\appendix
\restartappendixnumbering 

\section{Extended Morphological Models} \label{app:sec:morphs}

Three morphological models (Gaussian, Inverse power law, Diffusion) are tested for the modeling of the extended source HAWC~J1908+0618, which is a counterpart of MGRO~J1908+06. The mentioned morphological models were also tested for the modeling of MGRO~J1908+06 in \citet{Abeysekara:2018qtj}. Detailed implementation of the models can be found in \textit{astromodels}~\footnote{\textit{astromodels}; \url{https://github.com/threeML/astromodels}}. Astromodels is a Python package containing source definitions that is used with 3ML. Using the data in our ROI, we select the best model based on the relative $\Delta{\rm BIC}$~(Bayesian information criterion) obtained from the likelihood fits \citep{BIC}.

The Gaussian morphological model is described using
\begin{equation} \label{eq:gaussian}
\frac{\mathrm{d}N}{\mathrm{d}\Omega}=\frac{1}{2\pi\sigma^2}\exp{\left(-\frac{\theta^2}{2\sigma^2}\right)},
\end{equation}
where $\theta$ is angular distance and $\sigma$ is Gaussian width that represents the size of the fitted extended source in degrees. 

The inverse power law morphological model is given by
\begin{equation} \label{eq:powerlaw_morph}
\frac{\mathrm{d}N}{\mathrm{d}\Omega}= \left[\pi r_{\mathrm{min}}^{2+\alpha}+\frac{2\pi}{\left(2+\alpha\right)}\left(r_{\mathrm{max}}^{2+\alpha} - r_{\mathrm{min}}^{2+\alpha}\right)\right]^{-1} r^{\alpha},
\end{equation}
where $r_{\mathrm{min}} \leq \theta < r_{\mathrm{max}}$. 
$\Delta{\rm BIC}$ for the inverse power law model is found to be $+159$.

The diffusion morphological model describes $\gamma$-ray emission by inverse Compton scattering of diffused electrons and positrons as
\begin{equation} \label{eq:diffusion}
\frac{\mathrm{d}N}{\mathrm{d}\Omega}= \frac{1.22}{\pi^{3/2}\theta_{d}\left(\theta+0.06\theta_{d}\right)}\exp\left({-\frac{\theta^{2}}{\theta_{d}^{2}}}\right),
\end{equation} where $\theta_{d}$ is diffusion radius, which is the only free parameter used for the model \citep{HAWCGeminga}.
For the diffusion model, we have adopted the fixed parameters of \citet{HAWCJ1908} used to model MGRO~J1908+06, magnetic field of $3~\mu \mathrm{G}$ and diffusion coefficient spectral index of $1/3$.
$\Delta{\rm BIC}$ for the diffusion model compared to the Gaussian model is $-50~(={\rm BIC_{Diffusion}}-{\rm BIC_{Gaussian}})$. Overall, the diffusion model best describes HAWC~J1908+0618.

\section{Spectral Models} \label{app:sec:spectra}

Two spectral models (cutoff power-law and log parabola; COPL and LP) are tested for the found sources. The log parabola model is given by
\begin{equation} \label{eq:logp}   \frac{\mathrm{d}N}{\mathrm{d}E_{\gamma}} = K\left(\frac{E_{\gamma}}{E_{\mathrm{piv}}}\right)^{\alpha-\beta\log{\left(E_{\gamma}/E_{\mathrm{piv}}\right)}}, 
\end{equation} 
where $\beta$ is spectral curvature. 

A cutoff power-law model is given by 
\begin{equation} \label{eq:ecpl}   \frac{\mathrm{d}N}{\mathrm{d}E_{\gamma}} = K\left(\frac{E_{\gamma}}{E_{\mathrm{piv}}}\right)^{\alpha}\exp{\left(-\frac{E_{\gamma}}{E_{\mathrm{cutoff}}}\right)},
\end{equation} 
where $E_{\mathrm{cutoff}}$ is cutoff energy. The spectrum decreases exponentially when the energy is greater than the cutoff energy. Only the spectrum of HAWC~J1908+0618 is found to favour the LP model by $\Delta{\rm TS}_{\rm LP}=123$. 
COPL and LP do not have distinct preferences over the simple power law model for other sources, including the two SS~433 jet lobes.

\section{Energy estimator bins} \label{app:sec:energy_bins}

The values of flux points and upper limits presented in Figure~\ref{fig:hawc_spectra} are tabulated in Table~\ref{table:energybins}. The table also contains the energy range corresponding to each bin for the east and west SS~433 jet lobes.

\begin{table*}[ht!]
\centering
\begin{tabular}{c |cc| cc| cc| cc}
\multicolumn{1}{l}{\bf Bins} &
\multicolumn{2}{c}{\bf Median [TeV]} &
\multicolumn{2}{c}{\bf Lower Bound [TeV]} &
\multicolumn{2}{c}{\bf Upper Bound [TeV]} &
\multicolumn{2}{c}{\bf $E^2\text{d}N/\text{d}E$ [$10^{-13}$ TeV cm$^{-2}$ s$^{-1}$]}\\
\multicolumn{1}{c}{} &
\multicolumn{1}{c}{\bf west} &
\multicolumn{1}{c}{\bf east} &
\multicolumn{1}{c}{\bf west} &
\multicolumn{1}{c}{\bf east} &
\multicolumn{1}{c}{\bf west} &
\multicolumn{1}{c}{\bf east} &
\multicolumn{1}{c}{\bf west} &
\multicolumn{1}{c}{\bf east}\\
\hline\hline
c & 1.1 & 1.1 & 0.6 & 0.6 & 1.9 & 1.8 & $48.6^{*}$ & $2210^{*}$\\
d & 2.2 & 2.1 & 1.3 & 1.3 & 3.3 & 3.2 & $8.97^{*}$ & $9.01^{*}$\\
e & 3.8 & 3.8 & 2.6 & 2.5 & 5.5 & 5.4 & $5.18^{*}$ & $10.7^{*}$\\
f & 6.8 & 6.8 & 4.7 & 4.7 & 9.6 & 9.5 & $2.65^{+0.99}_{-0.69}$ & $2.47^{+0.92}_{-0.66}$\\
g & 12.4 & 12.3 & 9.0 & 8.9 & 16.7 & 16.6 & $1.95^{+0.73}_{-0.55}$ & $2.07^{+0.65}_{-0.51}$\\
h & 22.1 & 21.9 & 16.4 & 16.2 & 29.0 & 28.8 & $1.33^{+0.58}_{-0.40}$ & $1.27^{+0.52}_{-0.37}$\\
i & 39.0 & 38.7 & 29.5 & 29.3 & 50.9 & 50.5 & $2.12^{+0.57}_{-0.44}$ & $1.11^{+0.49}_{-0.34}$\\
j & 69.4 & 68.9 & 53.4 & 53.0 & 89.7 & 89.1 & $1.46^{*}$ & $1.16^{*}$\\
k & 124.9 & 124.2 & 96.9 & 96.3 & 160.1 & 159.1 & $0.851^{+0.655}_{-0.374}$ & $3.75^{*}$\\
l & 218.7 & 217.2 & 169.4 & 168.3 & 287.5 & 285.4 & $425^{*}$ & $217^{*}$\\
\hline
\end{tabular}
\caption{\label{table:energybins} Energy bins and their corresponding median, lower bound (16\% quantile), and upper bound (84\% quantile) energies in TeV. $E^2\text{d}N/\text{d}E$ represents the best-fit flux while $^{*}$ indicates upper limits. The upper limits are 90\% confidence level.}
\end{table*}

\bibliography{main}{}

\begin{thebibliography}{}
\expandafter\ifx\csname natexlab\endcsname\relax\def\natexlab#1{#1}\fi
\providecommand{\url}[1]{\href{#1}{#1}}
\providecommand{\dodoi}[1]{doi:~\href{http://doi.org/#1}{\nolinkurl{#1}}}
\providecommand{\doeprint}[1]{\href{http://ascl.net/#1}{\nolinkurl{http://ascl.net/#1}}}
\providecommand{\doarXiv}[1]{\href{https://arxiv.org/abs/#1}{\nolinkurl{https://arxiv.org/abs/#1}}}

\bibitem[{Abdo {et~al.}(2007)Abdo, Allen, Berley, Casanova, Chen, Coyne, Dingus, Ellsworth, Fleysher, Fleysher, Gonzalez, Goodman, Hays, Hoffman, Hopper, Hüntemeyer, Kolterman, Lansdell, Linnemann, McEnery, Mincer, Nemethy, Noyes, Ryan, Parkinson, Shoup, Sinnis, Smith, Sullivan, Vasileiou, Walker, Williams, Xu, \& Yodh}]{Abdo_2007}
Abdo, A.~A., Allen, B., Berley, D., {et~al.} 2007, The Astrophysical Journal, 664, L91, \dodoi{10.1086/520717}

\bibitem[{{Abeysekara} {et~al.}(2017){Abeysekara}, {Albert}, {Alfaro}, {Alvarez}, {{\'A}lvarez}, {Arceo}, {Arteaga-Vel{\'a}zquez}, {Ayala Solares}, {Barber}, {Bautista-Elivar}, {Becerril}, {Belmont-Moreno}, {BenZvi}, {Berley}, {Braun}, {Brisbois}, {Caballero-Mora}, {Capistr{\'a}n}, {Carrami{\~n}ana}, {Casanova}, {Castillo}, {Cotti}, {Cotzomi}, {Couti{\~n}o de Le{\'o}n}, {de la Fuente}, {De Le{\'o}n}, {DeYoung}, {Dingus}, {DuVernois}, {D{\'\i}az-V{\'e}lez}, {Ellsworth}, {Fiorino}, {Fraija}, {Garc{\'\i}a-Gonz{\'a}lez}, {Gerhardt}, {Gonz{\'a}lez Mun{\"o}z}, {Gonz{\'a}lez}, {Goodman}, {Hampel-Arias}, {Harding}, {Hernandez}, {Hernandez-Almada}, {Hinton}, {Hui}, {H{\"u}ntemeyer}, {Iriarte}, {Jardin-Blicq}, {Joshi}, {Kaufmann}, {Kieda}, {Lara}, {Lauer}, {Lee}, {Lennarz}, {Le{\'o}n Vargas}, {Linnemann}, {Longinotti}, {Raya}, {Luna-Garc{\'\i}a}, {L{\'o}pez-Coto}, {Malone}, {Marinelli}, {Martinez}, {Martinez-Castellanos}, {Mart{\'\i}nez-Castro}, {Mart{\'\i}nez-Huerta}, {Matthews}, {Miranda-Romagnoli}, {Moreno},
  {Mostaf{\'a}}, {Nellen}, {Newbold}, {Nisa}, {Noriega-Papaqui}, {Pelayo}, {Pretz}, {P{\'e}rez-P{\'e}rez}, {Ren}, {Rho}, {Rivi{\`e}re}, {Rosa-Gonz{\'a}lez}, {Rosenberg}, {Ruiz-Velasco}, {Salazar}, {Salesa Greus}, {Sandoval}, {Schneider}, {Schoorlemmer}, {Sinnis}, {Smith}, {Springer}, {Surajbali}, {Taboada}, {Tibolla}, {Tollefson}, {Torres}, {Ukwatta}, {Villase{\~n}or}, {Weisgarber}, {Westerhoff}, {Wisher}, {Wood}, {Yapici}, {Yodh}, {Younk}, {Zepeda}, \& {Zhou}}]{2017ApJ...843...39A}
{Abeysekara}, A.~U., {Albert}, A., {Alfaro}, R., {et~al.} 2017, \apj, 843, 39, \dodoi{10.3847/1538-4357/aa7555}

\bibitem[{Abeysekara {et~al.}(2017)Abeysekara, Albert, Alfaro, Alvarez, Álvarez, Arceo, Arteaga-Velázquez, Solares, Barber, Baughman, Bautista-Elivar, Gonzalez, Becerril, Belmont-Moreno, BenZvi, Berley, Bernal, Braun, Brisbois, Caballero-Mora, Capistrán, Carramiñana, Casanova, Castillo, Cotti, Cotzomi, de~León, de~la Fuente, León, Hernandez, Dingus, DuVernois, Díaz-Vélez, Ellsworth, Engel, Fiorino, Fraija, García-González, Garfias, Gerhardt, Muñoz, González, Goodman, Hampel-Arias, Harding, Hernandez, Hernandez-Almada, Hinton, Hui, Hüntemeyer, Iriarte, Jardin-Blicq, Joshi, Kaufmann, Kieda, Lara, Lauer, Lee, Lennarz, Vargas, Linnemann, Longinotti, Raya, Luna-García, López-Coto, Malone, Marinelli, Martinez, Martinez-Castellanos, Martínez-Castro, Martínez-Huerta, Matthews, Miranda-Romagnoli, Moreno, Mostafá, Nellen, Newbold, Nisa, Noriega-Papaqui, Pelayo, Pretz, Pérez-Pérez, Ren, Rho, Rivière, Rosa-González, Rosenberg, Ruiz-Velasco, Salazar, Greus, Sandoval, Schneider, Schoorlemmer, Sinnis,
  Smith, Springer, Surajbali, Taboada, Tibolla, Tollefson, Torres, Ukwatta, Vianello, Villaseñor, Weisgarber, Westerhoff, Wisher, Wood, Yapici, Younk, Zepeda, \& Zhou}]{2HWC}
Abeysekara, A.~U., Albert, A., Alfaro, R., {et~al.} 2017, The Astrophysical Journal, 843, 40, \dodoi{10.3847/1538-4357/aa7556}

\bibitem[{{Abeysekara} {et~al.}(2017){Abeysekara}, {Albert}, {Alfaro}, {Alvarez}, {{\'A}lvarez}, {Arceo}, {Arteaga-Vel{\'a}zquez}, {Avila Rojas}, {Ayala Solares}, {Barber}, {Bautista-Elivar}, {Becerril}, {Belmont-Moreno}, {BenZvi}, {Berley}, {Bernal}, {Braun}, {Brisbois}, {Caballero-Mora}, {Capistr{\'a}n}, {Carrami{\~n}ana}, {Casanova}, {Castillo}, {Cotti}, {Cotzomi}, {Couti{\~n}o de Le{\'o}n}, {De Le{\'o}n}, {De la Fuente}, {Dingus}, {DuVernois}, {D{\'\i}az-V{\'e}lez}, {Ellsworth}, {Engel}, {Enr{\'\i}quez-Rivera}, {Fiorino}, {Fraija}, {Garc{\'\i}a-Gonz{\'a}lez}, {Garfias}, {Gerhardt}, {Gonz{\'a}lez Mu{\~n}oz}, {Gonz{\'a}lez}, {Goodman}, {Hampel-Arias}, {Harding}, {Hern{\'a}ndez}, {Hern{\'a}ndez-Almada}, {Hinton}, {Hona}, {Hui}, {H{\"u}ntemeyer}, {Iriarte}, {Jardin-Blicq}, {Joshi}, {Kaufmann}, {Kieda}, {Lara}, {Lauer}, {Lee}, {Lennarz}, {Vargas}, {Linnemann}, {Longinotti}, {Luis Raya}, {Luna-Garc{\'\i}a}, {L{\'o}pez-Coto}, {Malone}, {Marinelli}, {Martinez}, {Martinez-Castellanos}, {Mart{\'\i}nez-Castro},
  {Mart{\'\i}nez-Huerta}, {Matthews}, {Miranda-Romagnoli}, {Moreno}, {Mostaf{\'a}}, {Nellen}, {Newbold}, {Nisa}, {Noriega-Papaqui}, {Pelayo}, {Pretz}, {P{\'e}rez-P{\'e}rez}, {Ren}, {Rho}, {Rivi{\`e}re}, {Rosa-Gonz{\'a}lez}, {Rosenberg}, {Ruiz-Velasco}, {Salazar}, {Salesa Greus}, {Sandoval}, {Schneider}, {Schoorlemmer}, {Sinnis}, {Smith}, {Springer}, {Surajbali}, {Taboada}, {Tibolla}, {Tollefson}, {Torres}, {Ukwatta}, {Vianello}, {Weisgarber}, {Westerhoff}, {Wisher}, {Wood}, {Yapici}, {Yodh}, {Younk}, {Zepeda}, {Zhou}, {Guo}, {Hahn}, {Li}, \& {Zhang}}]{HAWCGeminga}
{Abeysekara}, A.~U., {Albert}, A., {Alfaro}, R., {et~al.} 2017, Science, 358, 911, \dodoi{10.1126/science.aan4880}

\bibitem[{Abeysekara {et~al.}(2018)}]{Abeysekara:2018qtj}
Abeysekara, A.~U., {et~al.} 2018, Nature, 562, 82, \dodoi{10.1038/s41586-018-0565-5}

\bibitem[{Abeysekara {et~al.}(2019)}]{Abeysekara:2019edl}
---. 2019, Astrophys. J., 881, 134, \dodoi{10.3847/1538-4357/ab2f7d}

\bibitem[{Abeysekara {et~al.}(2021)}]{HAL}
---. 2021, PoS, ICRC2021, 828, \dodoi{10.22323/1.395.0828}

\bibitem[{{Abeysekara} {et~al.}(2023){Abeysekara}, {Albert}, {Alfaro}, {Alvarez}, {{\'A}lvarez}, {Araya}, {Arteaga-Vel{\'a}zquez}, {Arunbabu}, {Avila Rojas}, {Ayala Solares}, {Babu}, {Barber}, {Becerril}, {Belmont-Moreno}, {BenZvi}, {Blanco}, {Braun}, {Brisbois}, {Caballero-Mora}, {Cabrera Mart{\'\i}nez}, {Capistr{\'a}n}, {Carrami{\~n}ana}, {Casanova}, {Castillo}, {Chaparro-Amaro}, {Cotti}, {Cotzomi}, {Couti{\~n}o de Le{\'o}n}, {de la Fuente}, {de Le{\'o}n}, {De Young}, {Hernandez}, {Dingus}, {DuVernois}, {Durocher}, {D{\'\i}az-V{\'e}lez}, {Ellsworth}, {Engel}, {Espinoza}, {Fan}, {Fang}, {Fick}, {Fleischhack}, {Flores}, {Fraija}, {Garc{\'\i}a-Gonz{\'a}lez}, {Garcia-Torales}, {Garfias}, {Giacinti}, {Goksu}, {Gonz{\'a}lez}, {Gonz{\'a}lez-Mu{\~n}oz}, {Goodman}, {Harding}, {Hernandez}, {Hernandez}, {Hinton}, {Hona}, {Huang}, {Hueyotl-Zahuantitla}, {Hui}, {Humensky}, {H{\"u}ntemeyer}, {Iriarte}, {Imran}, {Jardin-Blicq}, {Joshi}, {Kaufmann}, {Kieda}, {Kunde}, {Lara}, {Lauer}, {Lee}, {Lennarz}, {Vargas},
  {Linnemann}, {Longinotti}, {Luis-Raya}, {Lundeen}, {Malone}, {Marandon}, {Marinelli}, {Martinez}, {Mart{\'\i}nez-Castellanos}, {Mart{\'\i}nez-Castro}, {Mart{\'\i}nez-Huerta}, {Matthews}, {Miranda-Romagnoli}, {Montaruli}, {Morales-Soto}, {Moreno}, {Mostaf{\'a}}, {Nayerhoda}, {Nellen}, {Newbold}, {Nisa}, {Noriega-Papaqui}, {Oceguera-Becerra}, {Olivera-Nieto}, {Omodei}, {Peisker}, {P{\'e}rez Araujo}, {P{\'e}rez-P{\'e}rez}, {Ponce}, {Pretz}, {Rho}, {Rosa-Gonz{\'a}lez}, {Ruiz-Velasco}, {Salazar}, {Salazar-Gallegos}, {Salesa Greus}, {Sandoval}, {Schneider}, {Schoorlemmer}, {Serna-Franco}, {Sinnis}, {Smith}, {Son}, {Sparks Woodle}, {Springer}, {Taboada}, {Tepe}, {Tibolla}, {Tollefson}, {Torres}, {Torres-Escobedo}, {Turner}, {Ure{\~n}a-Mena}, {Ukwatta}, {Varela}, {Vargas-Maga{\~n}a}, {Villase{\~n}or}, {Wang}, {Watson}, {Werner}, {Westerhoff}, {Willox}, {Wisher}, {Wood}, {Yodh}, {Zaborov}, {Zepeda}, {Zhou}, \& {HAWC Collaboration}}]{HAWCDetector}
{Abeysekara}, A.~U., {Albert}, A., {Alfaro}, R., {et~al.} 2023, Nuclear Instruments and Methods in Physics Research A, 1052, 168253, \dodoi{10.1016/j.nima.2023.168253}

\bibitem[{Albert {et~al.}(2024)Albert, Alfaro, Alvarez, Andrés, Arteaga-Velázquez, Rojas, Solares, Babu, Belmont-Moreno, Caballero-Mora, Capistrán, Carramiñana, Casanova, Cotti, Cotzomi, de~León, la~Fuente, de~León, Depaoli, Lalla, Hernandez, Dingus, DuVernois, Engel, Ergin, Espinoza, Fan, Fang, Fraija, Fraija, García-González, Garfias, Goksu, González, Goodman, Groetsch, Harding, Hernández-Cadena, Herzog, Hinton, Huang, Hueyotl-Zahuantitla, Hüntemeyer, Iriarte, Kaufmann, Lara, Lee, Vargas, Linnemann, Longinotti, Luis-Raya, Malone, Martínez-Castro, Matthews, Miranda-Romagnoli, Montes, Moreno, Mostafá, Nellen, Nisa, Noriega-Papaqui, Olivera-Nieto, Omodei, Osorio, Araujo, Pérez-Pérez, Rho, Rosa-González, Ruiz-Velasco, Salazar, Salazar-Gallegos, Sandoval, Schneider, Schwefer, Serna-Franco, Smith, Son, Springer, Tibolla, Tollefson, Torres, Torres-Escobedo, Turner, Ureña-Mena, Varela, Wang, Watson, Whitaker, Willox, Wu, Yu, Yun-Cárcamo, \& Zhou}]{Pass5Crab}
Albert, A.~., Alfaro, R., Alvarez, C., {et~al.} 2024, Performance of the HAWC Observatory and TeV Gamma-Ray Measurements of the Crab Nebula with Improved Extensive Air Shower Reconstruction Algorithms.
\newblock \doarXiv{2405.06050}

\bibitem[{{Albert} {et~al.}(2022){Albert}, {Alfaro}, {Alvarez}, {{\'A}lvarez}, {Angeles Camacho}, {Arteaga-Vel{\'a}zquez}, {Avila Rojas}, {Solares}, {Babu}, {Belmont-Moreno}, {Brisbois}, {Caballero-Mora}, {Capistr{\'a}n}, {Carrami{\~n}ana}, {Casanova}, {Cotti}, {Cotzomi}, {Couti{\~n}o de Le{\'o}n}, {De la Fuente}, {de Le{\'o}n}, {Diaz Hernandez}, {Dingus}, {DuVernois}, {Durocher}, {D{\'\i}az-V{\'e}lez}, {Engel}, {Espinoza}, {Fan}, {Fang}, {Fern{\'a}ndez Alonso}, {Fraija}, {Garcia}, {Garc{\'\i}a-Gonz{\'a}lez}, {Garfias}, {Giacinti}, {Goksu}, {Gonz{\'a}lez}, {Goodman}, {Harding}, {Hinton}, {Hona}, {Huang}, {Hueyotl-Zahuantitla}, {H{\"u}ntemeyer}, {Iriarte}, {Jardin-Blicq}, {Joshi}, {Kaufmann}, {Kieda}, {Lee}, {Lee}, {Vargas}, {Linnemann}, {Longinotti}, {Luis-Raya}, {Malone}, {Marandon}, {Martinez}, {Mart{\'\i}nez-Castro}, {Matthews}, {Miranda-Romagnoli}, {Morales-Soto}, {Moreno}, {Mostaf{\'a}}, {Nayerhoda}, {Nellen}, {Newbold}, {Nisa}, {Noriega-Papaqui}, {Olivera-Nieto}, {Omodei}, {Peisker}, {P{\'e}rez Araujo},
  {P{\'e}rez-P{\'e}rez}, {Rho}, {Rosa-Gonz{\'a}lez}, {Salazar}, {Salesa Greus}, {Sandoval}, {Schneider}, {Schoorlemmer}, {Serna-Franco}, {Smith}, {Son}, {Springer}, {Tibolla}, {Tollefson}, {Torres}, {Torres-Escobedo}, {Turner}, {Ure{\~n}a-Mena}, {Villase{\~n}or}, {Wang}, {Watson}, {Willox}, {Zepeda}, {Zhou}, {Breuhaus}, {Li}, {Zhang}, \& {HAWC Collaboration}}]{HAWCJ1908}
{Albert}, A., {Alfaro}, R., {Alvarez}, C., {et~al.} 2022, \apj, 928, 116, \dodoi{10.3847/1538-4357/ac56e5}

\bibitem[{Albert {et~al.}(2023)}]{4HWCCatalog}
Albert, A., {et~al.} 2023, PoS, ICRC2023, 759, \dodoi{10.22323/1.444.0759}

\bibitem[{{Bartoli} {et~al.}(2013){Bartoli}, {Bernardini}, {Bi}, {Bolognino}, {Branchini}, {Budano}, {Calabrese Melcarne}, {Camarri}, {Cao}, {Cardarelli}, {Catalanotti}, {Chen}, {Chen}, {Chen}, {Creti}, {Cui}, {Dai}, {D'Amone}, {Danzengluobu}, {De Mitri}, {D'Ettorre Piazzoli}, {Di Girolamo}, {Ding}, {Di Sciascio}, {Feng}, {Feng}, {Feng}, {Gou}, {Guo}, {He}, {Hu}, {Hu}, {Huang}, {Iacovacci}, {Iuppa}, {Jia}, {Labaciren}, {Li}, {Li}, {Li}, {Liguori}, {Liu}, {Liu}, {Liu}, {Liu}, {Lu}, {Ma}, {Ma}, {Mancarella}, {Mari}, {Marsella}, {Martello}, {Mastroianni}, {Montini}, {Ning}, {Panareo}, {Panico}, {Perrone}, {Pistilli}, {Ruggieri}, {Salvini}, {Santonico}, {Sbano}, {Shen}, {Sheng}, {Shi}, {Surdo}, {Tan}, {Vallania}, {Vernetto}, {Vigorito}, {Wang}, {Wang}, {Wu}, {Wu}, {Xu}, {Xue}, {Yang}, {Yang}, {Yao}, {Yuan}, {Zha}, {Zhang}, {Zhang}, {Zhang}, {Zhang}, {Zhang}, {Zhang}, {Zhang}, {Zhao}, {Zhaxiciren}, {Zhaxisangzhu}, {Zhou}, {Zhu}, {Zhu}, {Zizzi}, \& {ARGO-YBJ Collaboration}}]{2013ApJ...779...27B}
{Bartoli}, B., {Bernardini}, P., {Bi}, X.~J., {et~al.} 2013, \apj, 779, 27, \dodoi{10.1088/0004-637X/779/1/27}

\bibitem[{{Brinkmann} {et~al.}(1996){Brinkmann}, {Aschenbach}, \& {Kawai}}]{1996A&A...312..306B}
{Brinkmann}, W., {Aschenbach}, B., \& {Kawai}, N. 1996, \aap, 312, 306

\bibitem[{{Brinkmann, W.} {et~al.}(2007){Brinkmann, W.}, {Pratt, G. W.}, {Rohr, S.}, {Kawai, N.}, \& {Burwitz, V.}}]{refId0}
{Brinkmann, W.}, {Pratt, G. W.}, {Rohr, S.}, {Kawai, N.}, \& {Burwitz, V.} 2007, A\&A, 463, 611, \dodoi{10.1051/0004-6361:20065570}

\bibitem[{{Cao} {et~al.}(2021){Cao}, {Aharonian}, {An}, {Axikegu}, {Bai}, {Bao}, {Bastieri}, {Bi}, {Bi}, {Cai}, {Cai}, {Cao}, {Chang}, {Chang}, {Chang}, {Chen}, {Chen}, {Chen}, {Chen}, {Chen}, {Chen}, {Chen}, {Chen}, {Chen}, {Chen}, {Chen}, {Chen}, {Chen}, {Cheng}, {Cheng}, {Cui}, {Cui}, {Cui}, {Dai}, {Dai}, {Dai}, {Danzengluobu}, {della Volpe}, {D'Ettorre Piazzoli}, {Dong}, {Fan}, {Fan}, {Fan}, {Fang}, {Fang}, {Feng}, {Feng}, {Feng}, {Feng}, {Gao}, {Gao}, {Gao}, {Gao}, {Ge}, {Geng}, {Gong}, {Gou}, {Gu}, {Guo}, {Guo}, {Guo}, {Guo}, {Han}, {He}, {He}, {He}, {He}, {He}, {He}, {Heller}, {Hor}, {Hou}, {Hou}, {Hu}, {Hu}, {Hu}, {Hu}, {Huang}, {Huang}, {Huang}, {Huang}, {Huang}, {Ji}, {Ji}, {Jia}, {Jiang}, {Jiang}, {Jin}, {Kuleshov}, {Levochkin}, {Li}, {Li}, {Li}, {Li}, {Li}, {Li}, {Li}, {Li}, {Li}, {Li}, {Li}, {Li}, {Li}, {Li}, {Li}, {Li}, {Li}, {Liang}, {Liang}, {Lin}, {Liu}, {Liu}, {Liu}, {Liu}, {Liu}, {Liu}, {Liu}, {Liu}, {Liu}, {Liu}, {Liu}, {Liu}, {Liu}, {Liu}, {Liu}, {Long}, {Lu}, {Lv}, {Ma}, {Ma}, {Ma},
  {Mao}, {Masood}, {Mitthumsiri}, {Montaruli}, {Nan}, {Pang}, {Pattarakijwanich}, {Pei}, {Qi}, {Ruffolo}, {Rulev}, {S{\'a}iz}, {Shao}, {Shchegolev}, {Sheng}, {Shi}, {Song}, {Stenkin}, {Stepanov}, {Sun}, {Sun}, {Sun}, {Tam}, {Tang}, {Tian}, {Wang}, {Wang}, {Wang}, {Wang}, {Wang}, {Wang}, {Wang}, {Wang}, {Wang}, {Wang}, {Wang}, {Wang}, {Wang}, {Wang}, {Wang}, {Wang}, {Wang}, {Wang}, {Wang}, {Wang}, {Wang}, {Wei}, {Wei}, {Wei}, {Wen}, {Wu}, {Wu}, {Wu}, {Wu}, {Wu}, {Xi}, {Xia}, {Xia}, {Xiang}, {Xiao}, {Xiao}, {Xin}, {Xin}, {Xing}, {Xu}, {Xu}, {Xue}, {Yan}, {Yang}, {Yang}, {Yang}, {Yang}, {Yang}, {Yang}, {Yang}, {Yao}, {Yao}, {Ye}, {Yin}, {Yin}, {You}, {You}, {Yu}, {Yuan}, {Zeng}, {Zeng}, {Zeng}, {Zeng}, {Zha}, {Zhai}, {Zhang}, {Zhang}, {Zhang}, {Zhang}, {Zhang}, {Zhang}, {Zhang}, {Zhang}, {Zhang}, {Zhang}, {Zhang}, {Zhang}, {Zhang}, {Zhang}, {Zhang}, {Zhang}, {Zhang}, {Zhang}, {Zhang}, {Zhao}, {Zhao}, {Zhao}, {Zhao}, {Zhao}, {Zheng}, {Zheng}, {Zhou}, {Zhou}, {Zhou}, {Zhou}, {Zhou}, {Zhou}, {Zhu}, {Zhu}, {Zhu},
  {Zhu}, \& {Zuo}}]{UHELHAASO}
{Cao}, Z., {Aharonian}, F.~A., {An}, Q., {et~al.} 2021, \nat, 594, 33, \dodoi{10.1038/s41586-021-03498-z}

\bibitem[{{Cao} {et~al.}(2024){Cao}, {Aharonian}, {An}, {Axikegu}, {Bai}, {Bao}, {Bastieri}, {Bi}, {Bi}, {Cai}, {Cao}, {Cao}, {Cao}, {Chang}, {Chang}, {Chen}, {Chen}, {Chen}, {Chen}, {Chen}, {Chen}, {Chen}, {Chen}, {Chen}, {Chen}, {Chen}, {Chen}, {Cheng}, {Cheng}, {Cui}, {Cui}, {Cui}, {Cui}, {Dai}, {Dai}, {Dai}, {Danzengluobu}, {Della Volpe}, {Dong}, {Duan}, {Fan}, {Fan}, {Fang}, {Fang}, {Feng}, {Feng}, {Feng}, {Feng}, {Feng}, {Gabici}, {Gao}, {Gao}, {Gao}, {Gao}, {Gao}, {Gao}, {Ge}, {Geng}, {Giacinti}, {Gong}, {Gou}, {Gu}, {Guo}, {Guo}, {Guo}, {Guo}, {Han}, {He}, {He}, {He}, {He}, {He}, {Heller}, {Hor}, {Hou}, {Hou}, {Hou}, {Hu}, {Hu}, {Hu}, {Huang}, {Huang}, {Huang}, {Huang}, {Huang}, {Huang}, {Huang}, {Ji}, {Jia}, {Jia}, {Jiang}, {Jiang}, {Jiang}, {Jin}, {Kang}, {Ke}, {Kuleshov}, {Kurinov}, {Li}, {Li}, {Li}, {Li}, {Li}, {Li}, {Li}, {Li}, {Li}, {Li}, {Li}, {Li}, {Li}, {Li}, {Li}, {Li}, {Li}, {Li}, {Li}, {Liang}, {Liang}, {Lin}, {Liu}, {Liu}, {Liu}, {Liu}, {Liu}, {Liu}, {Liu}, {Liu}, {Liu}, {Liu}, {Liu},
  {Liu}, {Liu}, {Liu}, {Lu}, {Luo}, {Lv}, {Ma}, {Ma}, {Ma}, {Mao}, {Min}, {Mitthumsiri}, {Mu}, {Nan}, {Neronov}, {Ou}, {Pang}, {Pattarakijwanich}, {Pei}, {Qi}, {Qi}, {Qiao}, {Qin}, {Ruffolo}, {S{\'a}iz}, {Semikoz}, {Shao}, {Shao}, {Shchegolev}, {Sheng}, {Shu}, {Song}, {Stenkin}, {Stepanov}, {Su}, {Sun}, {Sun}, {Sun}, {Tam}, {Tang}, {Tang}, {Tian}, {Wang}, {Wang}, {Wang}, {Wang}, {Wang}, {Wang}, {Wang}, {Wang}, {Wang}, {Wang}, {Wang}, {Wang}, {Wang}, {Wang}, {Wang}, {Wang}, {Wang}, {Wang}, {Wang}, {Wang}, {Wang}, {Wei}, {Wei}, {Wei}, {Wen}, {Wu}, {Wu}, {Wu}, {Wu}, {Wu}, {Xi}, {Xia}, {Xia}, {Xiang}, {Xiao}, {Xiao}, {Xin}, {Xin}, {Xing}, {Xiong}, {Xu}, {Xu}, {Xu}, {Xu}, {Xue}, {Yan}, {Yan}, {Yan}, {Yang}, {Yang}, {Yang}, {Yang}, {Yang}, {Yang}, {Yang}, {Yang}, {Yang}, {Yao}, {Yao}, {Ye}, {Yin}, {Yin}, {You}, {You}, {Yu}, {Yuan}, {Yue}, {Zeng}, {Zeng}, {Zeng}, {Zha}, {Zhang}, {Zhang}, {Zhang}, {Zhang}, {Zhang}, {Zhang}, {Zhang}, {Zhang}, {Zhang}, {Zhang}, {Zhang}, {Zhang}, {Zhang}, {Zhang}, {Zhang}, {Zhang},
  {Zhang}, {Zhang}, {Zhao}, {Zhao}, {Zhao}, {Zhao}, {Zhao}, {Zheng}, {Zhou}, {Zhou}, {Zhou}, {Zhou}, {Zhou}, {Zhou}, {Zhou}, {Zhu}, {Zhu}, {Zhu}, {Zhu}, {Zuo}, \& {(The Lhaaso Collaboration)}}]{1LHAASOCat}
{Cao}, Z., {Aharonian}, F., {An}, Q., {et~al.} 2024, \apjs, 271, 25, \dodoi{10.3847/1538-4365/acfd29}

\bibitem[{{Fabrika}(2004)}]{2004ASPRv..12....1F}
{Fabrika}, S. 2004, \apspr, 12, 1, \dodoi{10.48550/arXiv.astro-ph/0603390}

\bibitem[{Fang {et~al.}(2020)Fang, Charles, \& Blandford}]{Fang_2020}
Fang, K., Charles, E., \& Blandford, R.~D. 2020, The Astrophysical Journal Letters, 889, L5, \dodoi{10.3847/2041-8213/ab62b8}

\bibitem[{{Geldzahler} {et~al.}(1980){Geldzahler}, {Pauls}, \& {Salter}}]{1980A&A....84..237G}
{Geldzahler}, B.~J., {Pauls}, T., \& {Salter}, C.~J. 1980, \aap, 84, 237

\bibitem[{{H.~E.~S.~S. Collaboration} {et~al.}(2024){H.~E.~S.~S. Collaboration}, {Olivera-Nieto}, {Reville}, {Hinton}, \& {Tsirou}}]{HESS24}
{H.~E.~S.~S. Collaboration}, {Olivera-Nieto}, L., {Reville}, B., {Hinton}, J., \& {Tsirou}, M. 2024, Science, 383, 402, \dodoi{10.1126/science.adi2048}

\bibitem[{Kar(2017)}]{Kar:2017oe}
Kar, P. 2017, PoS, ICRC2017, 713, \dodoi{10.22323/1.301.0713}

\bibitem[{Kass \& Raftery(1995)}]{BIC}
Kass, R.~E., \& Raftery, A.~E. 1995, Journal of the American Statistical Association, 90, 773, \dodoi{10.1080/01621459.1995.10476572}

\bibitem[{Kimura {et~al.}(2020)Kimura, Murase, \& M\'esz\'aros}]{Kimura:2020acy}
Kimura, S.~S., Murase, K., \& M\'esz\'aros, P. 2020, Astrophys. J., 904, 188, \dodoi{10.3847/1538-4357/abbe00}

\bibitem[{Reynoso \& Carulli(2019)}]{Reynoso:2019vrp}
Reynoso, M.~M., \& Carulli, A.~M. 2019, Astropart. Phys., 109, 25, \dodoi{10.1016/j.astropartphys.2019.02.003}

\bibitem[{Safi-Harb \& {\"O}gelman(1997)}]{Safi-Harb_1997}
Safi-Harb, S., \& {\"O}gelman, H. 1997, The Astrophysical Journal, 483, 868, \dodoi{10.1086/304274}

\bibitem[{Safi-Harb \& Petre(1999)}]{Safi-Harb_1999}
Safi-Harb, S., \& Petre, R. 1999, The Astrophysical Journal, 512, 784, \dodoi{10.1086/306803}

\bibitem[{Safi-Harb {et~al.}(2022)Safi-Harb, Intyre, Zhang, Pope, Zhang, Saffold, Mori, Gotthelf, Aharonian, Band, Braun, Fang, Hailey, Nynka, \& Rho}]{Safi-Harb_2022}
Safi-Harb, S., Intyre, B.~M., Zhang, S., {et~al.} 2022, The Astrophysical Journal, 935, 163, \dodoi{10.3847/1538-4357/ac7c05}

\bibitem[{{Stephenson} \& {Sanduleak}(1977)}]{1977ApJS.33.459S}
{Stephenson}, C.~B., \& {Sanduleak}, N. 1977, \apjs, 33, 459, \dodoi{10.1086/190437}

\bibitem[{{Sudoh} {et~al.}(2020){Sudoh}, {Inoue}, \& {Khangulyan}}]{Sudoh:2019jup}
{Sudoh}, T., {Inoue}, Y., \& {Khangulyan}, D. 2020, \apj, 889, 146, \dodoi{10.3847/1538-4357/ab6442}

\bibitem[{Vianello {et~al.}(2015)Vianello, Lauer, Younk, Tibaldo, Burgess, Ayala, Harding, Hui, Omodei, \& Zhou}]{threeML}
Vianello, G., Lauer, R.~J., Younk, P., {et~al.} 2015, The Multi-Mission Maximum Likelihood framework (3ML),  arXiv, \dodoi{10.48550/ARXIV.1507.08343}

\bibitem[{{Wakely} \& {Horan}(2008)}]{TeVCat}
{Wakely}, S.~P., \& {Horan}, D. 2008, International Cosmic Ray Conference, 3, 1341

\bibitem[{Wilks(1938)}]{Wilks}
Wilks, S.~S. 1938, The Annals of Mathematical Statistics, 9, 60 , \dodoi{10.1214/aoms/1177732360}

\end{thebibliography}
\bibliographystyle{aasjournal}

\end{document}